\documentclass[5p]{elsarticle}

\usepackage{lineno,hyperref}
\modulolinenumbers[100]

\usepackage{graphicx}% Include figure files
\usepackage{dcolumn}% Align table columns on decimal point
\usepackage{bm}% bold math
\usepackage{placeins} % to confine floats to their sections using \FloatBarrier.
\usepackage{amsmath}
\newcommand{\mr}[1]{\mathrm{#1}}
\renewcommand{\exp}[1]{\ensuremath{\mathrm{e}^{#1}}} 
\newcommand{\ii}{\ensuremath{\mathrm{i}}}
\newcommand{\vect}[1]{\ensuremath{\mathbf{#1}}}
\newcommand{\vectsym}[1]{\ensuremath{\boldsymbol{#1}}}
\journal{Colloids and surfaces A: physicochemical and engineering aspects}

\usepackage{numcompress}
\biboptions{sort&compress}

\begin{document}

\begin{frontmatter}

\title{Orienting lipid-coated graphitic micro-particles in solution using AC electric fields: a new theoretical \emph{dual-ellipsoid} Laplace model for electro-orientation}

% Authors:
\author{J. Nguyen}

\author{Jonathan G. Underwood}
 
\author{I. Llorente Garc\'{i}a}

\ead{i.llorente-garcia@ucl.ac.uk}

\address{Department of Physics and Astronomy, University College London, Gower St., London WC1E 6BT, UK.}

\date{\today}

\begin{abstract}
Graphitic micro-particles are commonly coated with thin layers to generate stable aqueous dispersions for various applications. Such particles are technologically interesting as they can be manipulated with electric fields. Modeling the electrical manipulation of submerged layered micro-particles analytically or numerically is not straightforward. In particular, the generation of reliable quantitative torque predictions for electro-orientation experiments has been elusive. The traditional Laplace model approximates the coated particle by an ellipsoid with a confocal ellipsoidal layer and solves Laplace's equation to produce convenient analytical predictions. However, due to the non-uniformity of the layer thickness around the ellipsoid, this method can lead to incorrect torque predictions. Here we present a new theoretical \emph{dual-ellipsoid} Laplace model that corrects the effect of the non-uniform layer thickness by calculating two layered ellipsoids, each accounting for the correct layer thickness along each relevant direction for the torque. Our model describes the electro-orientation of submerged lipid-coated graphitic micro-particles in the presence of an alternating current (AC) electric field and is valid for ellipsoids with moderate aspect ratios and coated with thin shells. It is one of the first models to generate correct quantitative electric torque predictions. We present model results for the torque versus frequency and compare them to our measurements for lipid-coated highly ordered pyrolytic graphite (HOPG) micro-flakes in aqueous NaCl solution at MHz frequencies. The results show how the lipid shell changes the overall electrical properties of the micro-flakes so that the torque is low at low frequencies and increases at higher frequencies into the MHz regime. The torque depends critically on the lipid-shell thickness, the solution conductivity and the shape of the particle, all of which can be used as handles to control the response of the particles. Our model is useful to predict the frequencies at which electro-orientation can be observed in dilute dispersions and the reduction in torque caused by the shell.
\end{abstract}


\end{frontmatter}

\linenumbers

\section{\label{sec:Intro} Introduction}

The controlled orientation and manipulation of carbon-based micro- and nano-particles such as graphite/graphene platelets and carbon nanotubes submerged in aqueous solution is of relevance to a number of scientific and technological challenges. We are particularly interested in the manipulation of individual graphitic micro-flakes for the development of new single molecule probes for sensing biologically relevant torque \cite{magnetoElectrOrientHOPG}. Also of interest are applications to biological and chemical sensing \cite{grapheneChemicalSensors2012,grapheneBiosensors2011,grapheneChemicalBioSensorsReview2010} and controlled fluid mixing in microfluidic devices for lab-on-a-chip applications \cite{actuatingSoftMatterTorque2016}. Recently, the controlled orientation by means of AC electric fields of graphene-oxide micro-platelets in dilute dispersions was applied to the switching of electro-optic display devices \cite{grapheneOxideMonolayersElectrAlign2014}. At higher particle concentrations, controlled orientation can lead to novel applications in materials science, for example, for the synthesis of artificial materials with tailored anisotropic properties \cite{magnAlignGrapheneOxideGel,superparamagnCoatingForAlignmtScience2012,alignedGrapheneThermalMaterialVacuumFiltration2011}, for the improvement of batteries \cite{magnAlignmentGraphiteBatteries} or for the development of graphene-based liquid crystals \cite{magnAlignGrapheneOxideLC, grapheneGraphiteFlakeElectricAlign2013,grapheneMonolayersLCelectrAlign2015}. In dense dispersions, inter-particle interactions and/or excluded-volume effects can affect the efficiency of particle orientation with external electric/magnetic fields and therefore require careful consideration.

Graphitic micro-particles are strongly hydrophobic and aggregate in aqueous solution to form clumps. In order to prevent this aggregation, it is common to generate dispersions by coating the particles with a layer of amphiphilic surfactant or polymer molecules which adsorb non-covalently to the particle's surface \cite{surfactant}. Alternatively, amphiphilic lipid molecules can be used to coat the particles \cite{grapheneLipids5,grapheneLipids3,grapheneLipids6, magnetoElectrOrientHOPG}. Any such added layers modify the overall electrical properties of the micro-particles in solution, influencing their manipulation with electric fields, which typically need to be high-frequency (MHz) AC fields. A clear theoretical framework that specifically describes the behaviour in solution of coated graphitic micro-particles in terms of their electrical manipulation has not been provided before and is therefore highly desirable for the generation of predictions and future experimental advances. 

In this paper, we focus on the orientation by means of electric fields of individual micron-sized graphitic flakes in solution with a view to quantifying the torque on the micro-particles for applications to sensing torque and rotary motion. Graphitic micro-flakes are strongly conducting and polarisable along the direction of the graphene planes and therefore ideal for electro-orientation experiments. However, possibly owing to the reduction in torque at low frequencies for submerged coated micro-flakes, success in experiments has only been achieved recently for graphite/graphene flakes in solution employing AC electric fields \cite{grapheneGraphiteFlakeElectricAlign2013,grapheneOxideMonolayersElectrAlign2014,grapheneMonolayersLCelectrAlign2015,magnetoElectrOrientHOPG}. 

\begin{figure}[ht]
	\centering
  \includegraphics[width=0.9\columnwidth]{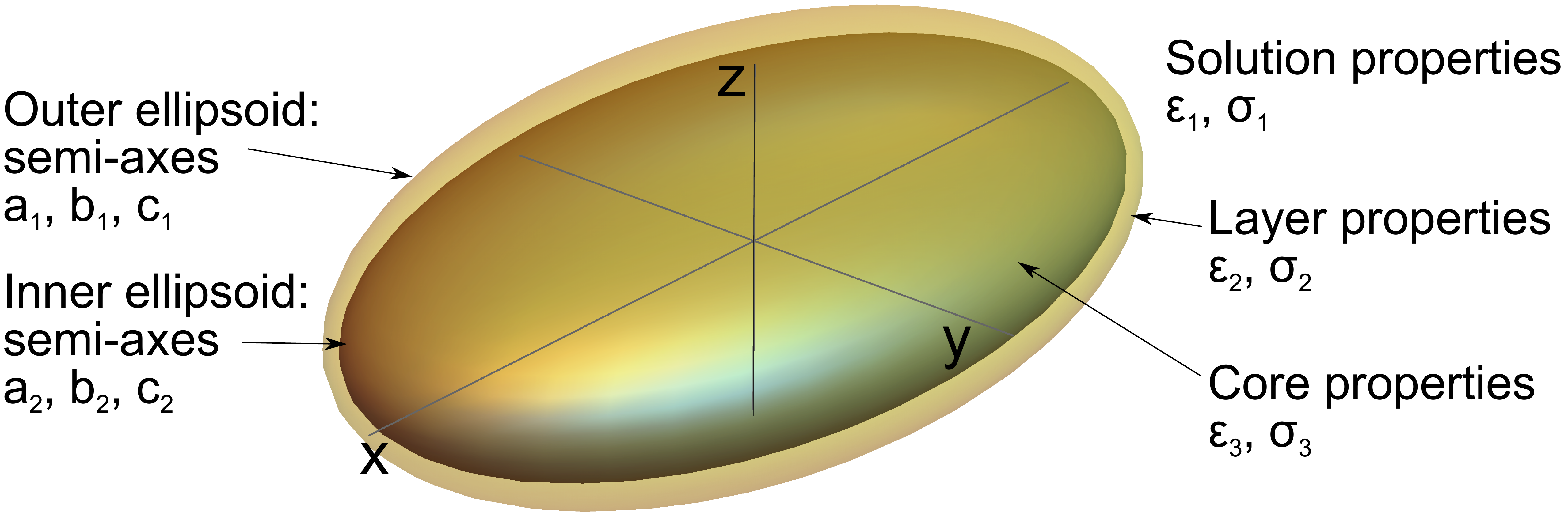}
  \caption{Schematic of layered ellipsoid in solution indicating nomenclature for properties of solution, lipid shell and HOPG core, and semi-axes for the inner and outer ellipsoids.}
  \label{fig:layeredEllipsoid}
\end{figure}

Modelling the electro-orientation of layered particles in solution by means of analytical or numerical methods is not easy. A widely used approach that generates convenient analytical predictions is the Laplace confocal ellipsoid (LCE) model. The solution of Laplace's equation can only be found analytically for confocal ellipsoids, the use of which implies that the layer thickness varies as a function of position on the ellipsoid surface (Fig. \ref{fig:layeredEllipsoid}). When a particle coated with a layer of uniform thickness is modeled as a confocal ellipsoid, the error introduced by the layer-thickness variation can be significant even for layers that are thin compared to the particle dimensions. Also, the model may not apply well to shelled particles with very large aspect ratios \cite{T12}. 
As the confocal layer is typically thinner along the longest ellipsoid dimension and thicker along the shortest particle dimension (compared to the desired uniform thickness) the model can substantially overestimate the torque by overestimating the effective polarization along the longest dimension and underestimating that along the shortest dimension. 

In this work, in order to correct for the effect of the non-uniform layer thickness while maintaining the use of convenient analytical expressions, we present a new theoretical \emph{dual-ellipsoid} Laplace model for the electro-orientation of submerged layered micro-particles. The model encompasses calculations for two different layered ellipsoids in order to account for the correct layer thickness along each relevant direction for the torque. The model is valid for ellipsoids with moderate aspect ratios with shells that are thin compared to the particle size. We apply our theory to the electro-orientation of lipid-coated highly ordered pyrolytic graphite (HOPG) micro-flakes in 20 mM NaCl aqueous solution and compare to experiments at frequencies $10-70\,\mathrm{MHz}$ \cite{magnetoElectrOrientHOPG} with good quantitative agreement. We highlight the most critical parameters that can be varied in experiments to modify the electric torque exerted on the micro-particles. Our results apply to individual micro-particles and to dilute dispersions of micro-particles in which particle-particle interactions can be neglected. For our micron-sized particles, the latter implies volume fractions $\phi \le 2.5 \,\mr{vol}\%$ or weight fractions $w \leq 1 \,\mr{wt}\%$ (see \ref{App:particleDistance} for more details on these considerations).

\section{\label{sec:lipidCoatedHOPG} Solubilized lipid-coated HOPG micro-particles}

The possibility of stably coating mono- and few-layer graphene sheets in aqueous solution with uniform phospholipid monolayers has been demonstrated both theoretically with molecular dynamics simulations \cite{grapheneLipids1,grapheneLipids1b}, and experimentally for both pristine graphene \cite{grapheneLipids5} and reduced graphene oxide \cite{grapheneLipids3,grapheneLipids6}. Lipid-coating of carbon nanotubes has also been shown \cite{lipidCNT1,lipidCNT2,lipidCNT3}. 

HOPG micro-flakes are stacks of graphene planes with very small mosaic spread. We have recently generated solubilized micron-sized HOPG particles in aqueous NaCl solution by coating them with multilayers of neutral POPC phospholipids \cite{magnetoElectrOrientHOPG}. On average, the micro-particles have a thickness of $1-2\,\mu \mathrm{m}$ and a lateral size $\sim 7\,\mu \mathrm{m} \times 3\,\mu \mathrm{m}$. The hydrophobic tails in the phospholipid molecules self-assemble onto the graphene planes on the HOPG surface via non-covalent hydrophobic interactions forming an initial lipid monolayer (as they do on mono- and few-layer graphene). A lipid multilayer can then self-assemble via adsorption of additional lipid bilayers to the first lipid monolayer. Atomic force microscopy (AFM) images show full lipid coverage of the HOPG particles with the external lipid multilayers forming a vesicle-like structure; AFM force-indentation traces suggest that the multilayer coating contains several lipid bilayers \cite{magnetoElectrOrientHOPG}.

\section{\label{sec:electric} Theory of AC electro-orientation of layered micro-particles in solution}

In this section, we first review existing theoretical models for layered particles in solution subjected to time-varying electric fields. We then summarise the LCE theory which is qualitatively correct and captures the main physics of the problem, e.g., the dependence of the torque on frequency. Following that, we evaluate the significance of the shell-thickness variation with position on the ellipsoid surface within the LCE model and explain our new \emph{dual-ellipsoid} Laplace model that accounts for the correct shell thickness along each dimension relevant for the electro-orienting torque. 

\subsection{\textbf{Existing theoretical models}} \label{sec:existingModels}

The theoretical description of the electrical properties of shelled particles in solution in the presence of alternating electric fields was first developed by Maxwell and Wagner for spherical particles \cite{Maxwell,Wagner}. The theory was later extended to ellipsoidal shelled particles \cite{T1,T15,AsamiEllipsoidShell,TBJonesBook} and also to multi-shelled submerged ellipsoids \cite{KakutaniEllipsoidShell}. This theory can be applied to various particle electro-manipulation experiments, such as electro-orientation (orientation using a linearly polarised AC electric field), electro-rotation (rotation in the presence of a rotating electric field), dielectrophoresis (transport in a spatially varying electric field), electrical impedance characterisation, etc. \cite{TBJonesBook}. 

Following the framework of the LCE model, Laplace's equation can be solved to obtain the electrostatic potential, $\Psi (r,t)$, in all regions of space. The solution for $\Psi (r,t)$ far outside the particle can be compared to that for a point dipole in order to extract an expression for the effective electric dipole moment of the particle using the so-called effective moment method \cite{Maxwell,T15,AsamiEllipsoidShell,BohrenHuffmanBook,TBJonesBook}. This effective moment represents the moment that a point dipole, located at the same position as the particle in suspension, would have in order to generate the same dipolar contribution to the electrostatic potential as the actual layered particle. This framework assumes that the electric field is uniform over the particle size and considers only dipolar terms, neglecting higher-order multipolar components and particle-particle interactions. The solution of Laplace's equation implies the application of appropriate boundary conditions to satisfy Maxwell's equations at all boundary surfaces (solution-shell and shell-core interfaces) and the fact that the applied external field must be recovered far away from the particle. Continuity of the electrical potential is applied, as well as continuity of the normal electrical displacement vector in its complex form, thereby incorporating the charge-continuity condition to account for the dynamic accumulation of surface charges at the boundary of two media with different conductivities. Hence, each interface (with different electrical properties at either side of the boundary) implies the possible accummulation of free charge at the boundary surface and the existence of so-called Maxwell-Wagner polarisation/relaxation effects, which are time- and frequency-dependent. Making use of the complex form for all permittivities and variables involved in the solution of Laplace's equation guarantees that Maxwell-Wagner effects at all interfaces are accounted for in the formulation \cite{TBJonesBook}. 

The model does not account for interfacial surface currents or for the effect of the electrical double layer (EDL) (a layer of ions, typically $\sim 1\,\mr{nm}$-thick, acquired by particles that display charged or polar groups on their surface in solution). Extended models that account for interfacial mobile surface charges can be found in the literature \cite{T9,T10,T11,TBJonesBook,T16}. Tangential surface currents can have an important contribution when the difference between the surface conductivity and the conductivity of the adjacent medium is large \cite{T9,T10,T11,T16} and their effect is only relevant at low frequencies ($<10^4\,\mr{Hz}$) \cite{T11,T16}. Various models have been derived for the specific case of the EDL that contributes to the Maxwell-Wagner polarization and can also harbour tangential field-driven ions (see \cite{T11} and references therein). In the case of AC electrical manipulation of nanometre-sized suspended particles or nano-scale biological molecules (e.g., proteins) the EDL can have a thickness comparable to or larger than the nano-particle size and its effect cannot be neglected \cite{electrDoubleLayers,BoundWaterEffects,proteinDEP,biowater}. Other authors have extended the LCE model to account for the possible electrical anisotropy of the shell (e.g., anisotropic lipid bilayer), however, the effect was found to be negligible for thin lipid-shell membranes \cite{T16}. 

Alternative models based on the AC response of resistor-capacitor (RC) electrical networks (electrical circuit equivalent to the suspended layered particle) have also been developed and applied mostly to the study of multi-particle suspensions (see \cite{T6} and references therein). Most often, agreement with experiments has been qualitative, although RC models accounting for particle geometry have produced results comparable to the LCE model \cite{T8,T6}. Refinements of these RC models have also considered layers with uniform thickness \cite{T12,T8}. 

All the above models ignore the possible contribution of time-varying magnetic fields that may arise in the presence of time-varying electric fields and surface currents, and consequently ignore induction effects. In general, the above theories have found a much easier application in the study of the effective electrical properties of many-particle suspensions \cite{AsamiEllipsoidShell,T6,T8}, whereas quantitative agreement has been elusive for single-particle measurements. This is the case particularly for electro-orientation, where quantitative torque measurements have been rarely reported and agreement with theory has been mostly qualitative. 

In this manuscript, we model the orientation of micron-sized lipid-coated HOPG flakes in NaCl aqueous solution. The flakes, that have low aspect ratios ($<6$) and are coated with very thin lipid shells (tens of nm thick), are modeled as layered ellipsoids in solution (Fig.\ref{fig:layeredEllipsoid}). We can neglect the electrical anisotropy of the thin lipid shell, as explained earlier \cite{T16}. We focus on modeling experiments in the MHz regime. At these frequencies, the contributions of EDL effects and possible surface currents can be neglected. Such effects could be possible in principle due to the possible adsorption of $\mr{Na}^+$ ions to the polar head groups of the POPC phospholipids in the lipid layer in NaCl solution \cite{T14}. However, these contributions are likely small for neutral lipids and would appear at frequencies $<10^4\,\mr{Hz}$ \cite{T11,T16}.

\subsection{\textbf{One-shell one-ellipsoid LCE model}} \label{sec:OneLayer}

We summarise here the one-shell standard LCE method for completion and to aid the following discussion and application. The theory derives from various references in different contexts, e.g., for particle suspensions and for different electro-kinetics experiments (not limited to electro-orientation) \cite{T15,AsamiEllipsoidShell,TBJonesBook,T3}. 

Modelling lipid-coated HOPG micro-flakes as submerged layered ellipsoids, we define a particle frame of reference with axes ($x$, $y$, $z$) fixed to the particle so that the $x$-$y$ plane corresponds to the graphene planes and $z$ is normal to these planes (Fig. \ref{fig:layeredEllipsoid}). The layered ellipsoid has an inner HOPG core and one confocal shell (lipid shell) that can be made out of several lipid bilayers. The inner ellipsoid has semi-axes $a_2$, $b_2$, $c_2$ along the $x$, $y$ and $z$ directions, respectively, while the outer ellipsoid has semi-axes $a_1$, $b_1$, $c_1$, along the same directions, with $a_1=(a_2^2+\delta)^{1/2}$, $b_1=(b_2^2+\delta)^{1/2}$ and $c_1=(c_2^2+\delta)^{1/2}$. $\delta$ is the parameter that defines the family of confocal ellipsoids. The actual thickness of the shell along each direction is given by $t_x=a_1-a_2$, $t_y=b_1-b_2$ and $t_z=c_1-c_2$.

The dielectric and conducting electrical properties of the core, shell and solution are considered by means of complex permittivities, $\epsilon$. We have, $\epsilon_1 = \varepsilon_1-\ii \sigma_1/(\omega \varepsilon_0)$ for the solution, $\epsilon_2 = \varepsilon_2-\ii \sigma_2/(\omega \varepsilon_0)$ for the shell, and we distinguish between in-plane ($\parallel$) and out-of-plane ($\perp$) properties for the anisotropic HOPG core, having $\epsilon_{3\parallel} = \varepsilon_{\mr{hopg},\parallel}-\ii \sigma_{\mr{hopg},\parallel}/(\omega \varepsilon_0)$ and $\epsilon_{3\perp} = \varepsilon_{\mr{hopg},\perp}-\ii \sigma_{\mr{hopg},\perp}/(\omega \varepsilon_0)$. In the previous expressions, $\varepsilon$ and $\sigma$ are the symbols for the static relative permittivity and conductivity, respectively, $\varepsilon_0$ is the permittivity of free space and $\omega = 2 \pi f$ is the angular frequency of the time-varying electric field with frequency $f$. 

In the presence of the externally applied AC electric field, $\vect{E}(t)$, the instantaneous electric torque acting on the submerged particle can be calculated as:
\begin{equation}\label{eqn:instantTorq} 
\vectsym{\mathcal{T}}\left(t\right)=\mr{Re}\left[\vect{p}_{\mr{eff}}(t) \right] \times \mr{Re}\left[\vect{E} (t) \right]\,,
\end{equation}
where $\vect{p}_{\mr{eff}}(t)$ is the effective electric dipole moment induced on the particle. We use complex notation so that $\vect{E}(t)=\vect{E}_0 \, \exp{\ii \omega t}$, where $\vect{E}_0$ is the time-independent complex field amplitude. Similarly, $\vect{p}_{\mr{eff}}(t)=\vect{p}_{\mr{eff,0}}\,\exp{\ii \omega t}$, where $\vect{p}_{\mr{eff,0}}$ is the complex vector amplitude of the induced electric dipole moment. The instantaneous torque consists of an average constant term and a time-varying term that oscillates at frequency $2\omega$ and that is quickly damped out by the dominant viscosity of the solution for micrometre-sized particles \cite{TBJonesBook}. Hence, only the time-averaged electric torque is relevant for our micro-particles, given by:
\begin{equation}\label{eqn:timeAvgTorq0} 
\left\langle \vectsym{\mathcal{T}} \right\rangle = \frac{1}{2}\,\mr{Re}\left[\vect{p}_{\mr{eff},0} \times \vect{E}^*_0 \right]\,,
\end{equation}
where $^*$ indicates complex conjugate.

The effective electric dipole moment, $\vect{p}_{\mr{eff}}(t)$, of the confocal layered ellipsoid in solution is obtained by solving Laplace's equation following the effective moment method outlined above \cite{Maxwell,T15,AsamiEllipsoidShell,TBJonesBook}. The vector components of $\vect{p}_{\mr{eff}}(t)$ are given by:
\begin{equation}\label{eqn:peff}
	p_{\mr{eff},k} (t) = V_1 \varepsilon_1 \varepsilon_0  K_k E_{k}(t)\, ,
\end{equation}
where $k=x,\,y\,,z$, $E_{k}(t)$ are the components of the external field, $V_1=\frac{4}{3}\pi a_1 b_1 c_1$ is the volume of the outer ellipsoid and $K_k$ are the generalized complex effective polarization factors. The latter depend on the electric field frequency, the particle geometry and the electrical properties of both particle and solution, and can be expressed as:
\begin{equation}\label{eqn:Kk}
	K_k = \frac{\epsilon'_{2k}-\epsilon_1}{\epsilon_1+(\epsilon'_{2k}-\epsilon_1)L_{1k}}\, .
\end{equation}
Here $L_{1k}$ are the geometrical, so-called depolarization factors \cite{Landau} of the outer ellipsoid [Eqn.(\ref{eqn:L1k}), \ref{App:depolarisationFactors}] and $\epsilon'_{2k}$ are the equivalent complex permittivities of the layered ellipsoid that are different along each axis (the shell introduces additional anisotropy \cite{TBJonesBook}) and are given by: 
\begin{equation}\label{eqn:equivalentPermittivity}
	\epsilon'_{2k} = 
		\epsilon_2 \cdot 
		\frac{\epsilon_2+\left(\epsilon_{3k}-\epsilon_2 \right) \left( L_{2k}-\nu L_{1k}+\nu \right)}
		{\epsilon_2+\left(\epsilon_{3k}-\epsilon_2 \right) \left( L_{2k}-\nu L_{1k} \right)} \, .
\end{equation}
$L_{2k}$ are the geometrical depolarization factors of the inner ellipsoid [Eqn.(\ref{eqn:L2k}), \ref{App:depolarisationFactors}] and $\nu=(a_2 b_2 c_2)/(a_1 b_1 c_1)$. Given that results can vary substantially depending on the particle geometry, we take the general approach of three different semi-axes and avoid oblate/prolate ellipsoid approximations. $\epsilon_{3k}$ are the components of the complex permittivity of the inner ellipsoid. We have accounted for the material anisotropy of the graphitic core by inserting an anisotropic $\epsilon_{3k}$ in Eqn.(\ref{eqn:equivalentPermittivity}). A similar approach was followed by Radu \textit{et al.} \cite{RetinalRods} to model the AC electro-orientation of retinal rods. For our HOPG particles we identify $\epsilon_{3x} = \epsilon_{3y} = \epsilon_{3\parallel}$ and $\epsilon_{3z} = \epsilon_{3\perp}$ for the in-plane and out-of-plane HOPG directions respectively, using the previously defined $\epsilon_{3\parallel}$ and $\epsilon_{3\perp}$. 

Now, following from Eqn.(\ref{eqn:timeAvgTorq0}), the time-averaged torque components can be calculated as:
{\setlength{\mathindent}{0cm}
	\begin{eqnarray}\label{eqn:timeAveragedTorqueComponents}
		\left\langle \mathcal{T} \right\rangle_\alpha & = & \frac{1}{2} \mr{Re}\left[ p_{\mr{eff},0 \beta} \, E^*_{0 \gamma} - 
															p_{\mr{eff},0 \gamma}\, E^*_{0 \beta} \right]  \nonumber \\ 
																									& = & \frac{1}{2} V_1 \varepsilon_1 \varepsilon_0 \mr{Re}\left[ K_\beta E_{0 \beta} E^*_{0 \gamma}- K_\gamma E_{0 \gamma} E^*_{0 \beta} \right]	\,,
	\end{eqnarray}
	}
with $\alpha,\,\beta,\,\gamma$ being cyclic coordinates (i.e., $\alpha=x$, $\beta=y$, $\gamma=z$ or $\alpha=y$, $\beta=z$, $\gamma=x$ or $\alpha=z$, $\beta=z$, $\gamma=y$). The above equation is valid in general for any electric field of the form $\vect{E}(t)=\vect{E}_0 \, \exp{\ii \omega t}$.

\begin{figure}[ht]
	\centering
  \includegraphics[width=0.8\columnwidth]{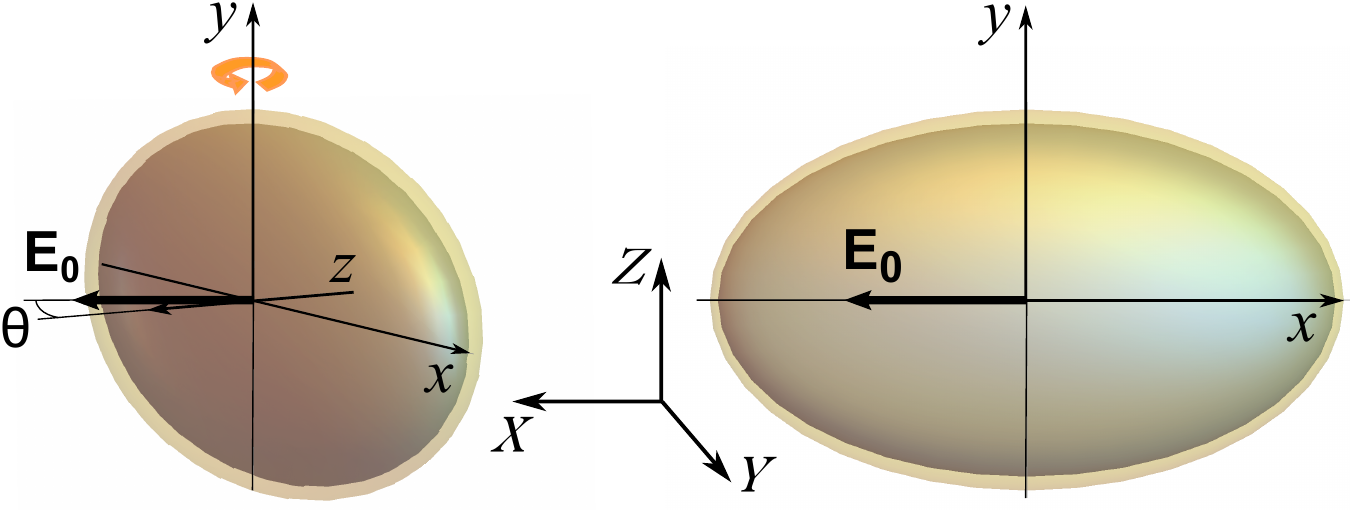}
  \caption{Electro-orientation of layered ellipsoid in a horizontal, linearly-polarized AC electric field that makes an angle $\theta$ to the direction perpendicular to the HOPG graphene planes.}
  \label{fig:rotation}
\end{figure}

We now define a fixed laboratory frame of reference with axes ($X$, $Y$, $Z$) (see Fig. \ref{fig:rotation}). In what follows, for the sake of simplicity and to compare to our experimental data, we apply the following assumptions: i) the micro-particle is pre-aligned with its graphene planes parallel to $Z$ (in experiments, we used a vertical magnetic field along $Z$ to force the HOPG micro-flakes to orient vertically \cite{magnetoElectrOrientHOPG}), and ii) the electric field is linearly polarized along the horizontal $X$ direction and makes an angle $\theta$ to the out-of-plane $z$ axis (particle frame of reference), as shown in Fig.\ref{fig:rotation}. We can then write $\vect{E}_0 = E_0 \left( \sin \theta\,\hat{\vect{x}} + \cos \theta\,\hat{\vect{z}} \right)$, where $E_0$ is the field amplitude.

The time-averaged torque exerted on the particle is around its in-plane direction $y$, and takes the form:
\begin{equation}\label{eqn:timeAveragedTorqueComponents2}
	\left\langle \mathcal{T} \right\rangle_{\parallel} = 
	\frac{1}{4} V_1 \varepsilon_1 \varepsilon_0 E_0^2 \sin(2\theta)  \mr{Re}\left[K_z-K_x\right] \, ,
\end{equation}
where $K_x$ and $K_z$ [see Eqn. (\ref{eqn:Kk})] are, respectively, the in-plane and out-of-plane effective polarization factors for the lipid-coated graphitic micro-particles. This torque makes the micro-flakes rotate until their graphene planes align parallel to the applied electric field direction, reaching stable rotational equilibrium when $\theta = \pm \pi/2$. Additionally, we define $\mathcal{T}_\mr{max}$ as the maximum amplitude of the time-averaged electric torque (at angles $\theta=\pm \pi/4$):
\begin{equation}\label{eqn:Tmax}
					\mathcal{T}_\mr{max}(\omega) = 
	\frac{1}{4} V_1 \varepsilon_1 \varepsilon_0 E_0^2 \mr{Re}\left[K_x(\omega)-K_z(\omega)\right] \, .
\end{equation}

%The time-averaged electric interaction energy is given by: 
%\begin{eqnarray}\label{eqn:Uelectr}
	%U_\mr{e} & = & -\int_{0}^{E_0} \left\langle \mr{Re}\left[\vect{p}_{\mr{eff}}(t)\right] \cdot \mr{Re}\left[\mr{d}\vect{E}(t)\right] 														\right\rangle  \nonumber \\ 
	    %& = & -\frac{V_2 \varepsilon_1 \varepsilon_0 E_0^2}{4} \mr{Re}\left[K_{\parallel} + \left(K_{\perp} - K_{\parallel}\right)\cos^2\theta \right] \,.
%\end{eqnarray} 

% Note that \cite{magnAlignmentGraphiteAstro} has incorrect expressions for the magnetic interaction energy. 

\subsection{\textbf{Variation with frequency}} \label{sec:versusFreq}

\begin{figure}[ht]
	\centering
  \includegraphics[width=0.73\columnwidth]{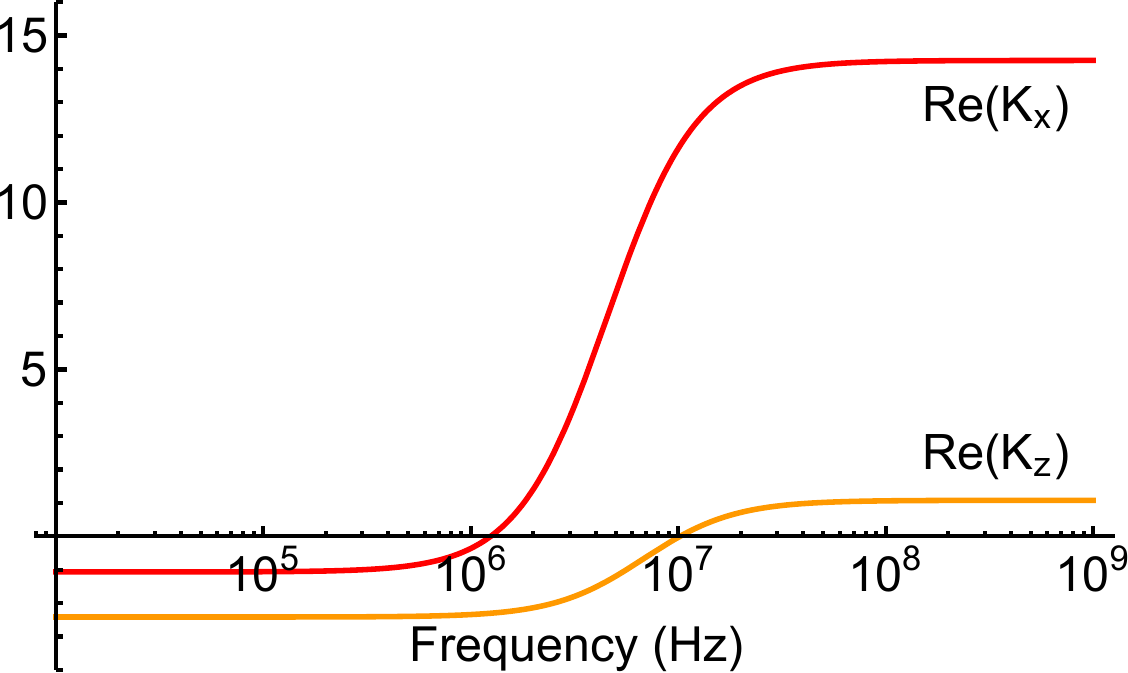}
  \caption{$\mr{Re}(K_x)$ and $\mr{Re}(K_z)$ versus frequency for a lipid-coated ellipsoidal HOPG particle calculated with the one-shell one-ellipsoid LCE model using a particle size $a_2=3\,\mu\mr{m}$, $b_2=0.83\,\mu\mr{m}$ and $c_2=0.5\,\mu\mr{m}$ and $\delta=(0.14\,\mu \mr{m})^2$ which results in a layer thickness $\sim10\,\mr{nm}$ on average ($t_x\approx3\,\mr{nm}$ and $t_z\approx18\,\mr{nm}$).}
  \label{fig:KxKzVsFreq}
\end{figure}

Figure \ref{fig:KxKzVsFreq} plots $\mr{Re}[K_x(\omega)]$ and $\mr{Re}[K_z(\omega)]$ as a function of the electric field frequency. Plots are shown for an ellipsoidal micro-particle with a lipid-layer thickness $\sim10\,\mr{nm}$, for the one-shell standard LCE model in section \ref{sec:OneLayer}. 

The variation with frequency of $\mr{Re}(K_x)$ and $\mr{Re}(K_z)$ arises as follows. Phospholipid bilayers, known to be highly electrically insulating, polarise in the presence of the electric field, giving rise to a large capacitive impedance that decreases with increasing frequency. As shown in Fig. \ref{fig:KxKzVsFreq}, the lipid coating insulates the micro-particle from the AC electric field at frequencies below $1-10\,\mr{MHz}$, reducing the effective induced electric dipole moment along each direction at these frequencies. We can model the lipid shell as a capacitor in parallel with a resistor ($\epsilon_2 = \varepsilon_2-\ii \sigma_2/(\omega \varepsilon_0)$). At low frequencies, the capacitive impedance of the shell (proportional to $1/(\omega C)$, where $C$ is the shell capacitance) is very large and the layered ellipsoid behaves like an insulating particle, giving rise to the observed low-frequency plateaus in the figure. In this regime, free surface charges accumulate at the solution-lipid interface (outer shell interface) according to the difference in conductivity between these two media. As the frequency increases, the frequency-dependent capacitive impedance of the lipid shell decreases and becomes first comparable to and then increasingly lower than its resistive impedance, until the lipid shell becomes electrically transparent at large frequencies (shell bridging). This capacitive decrease in impedance gives rise to increasing values of $\mr{Re}(K_x)$ and $\mr{Re}(K_z)$ with frequency that corresponds to the large dispersive features shown in the figure between $\sim1\,\mr{MHz}$ and $\sim10\,\mr{MHz}$. At high enough frequencies, full capacitive shell bridging takes place and the resistive, frequency-independent impedance of HOPG dominates, leading to the high-frequency constant plateaus. In this regime, polarisation depends on the balance of conductivity between the ionic solution and the HOPG core.

As a consequence, as shown in Fig. \ref{fig:KxKzVsFreq}, at low frequencies we have negative values of $\mr{Re}(K_x)$ and $\mr{Re}(K_z)$ (also known as regime of negative dielectrophoresis) while at high frequencies, we have positive values (positive dielectrophoresis). The direction of stable orientation of the layered particle is given by the axis $k$ with the largest value of $\mr{Re}(K_k)$ (including sign). This is because $-\mr{Re}(K_k)$ is a measure of the energy of the ellipsoid when oriented with its $k$ axis parallel to the electric field \cite{T1,magnetoElectrOrientHOPG,TBJonesBook}. Since $\mr{Re}(K_x)>\mr{Re}(K_z)$ at all frequencies, the lipid-coated HOPG micro-flakes stably align with their longest axis ($a_2$) along the field direction \cite{magnetoElectrOrientHOPG}. Only if there are cross-overs between $\mr{Re}(K_x)$ and $\mr{Re}(K_z)$, there can be changes in the axis of stable orientation \cite{T1,TBJonesBook}. This is not the case for the set of parameters explored for lipid-coated HOPG micro-flakes in our experiments. Changes in the stable orientation axis have been observed, e.g., for llama erythrocytes (micron-sized ellipsoidal cells) \cite{T3}.

\subsection{\textbf{Dependence on shell thickness}} \label{sec:layerThickness}

\begin{figure}[ht]
	\centering
  \includegraphics[width=0.77\columnwidth]{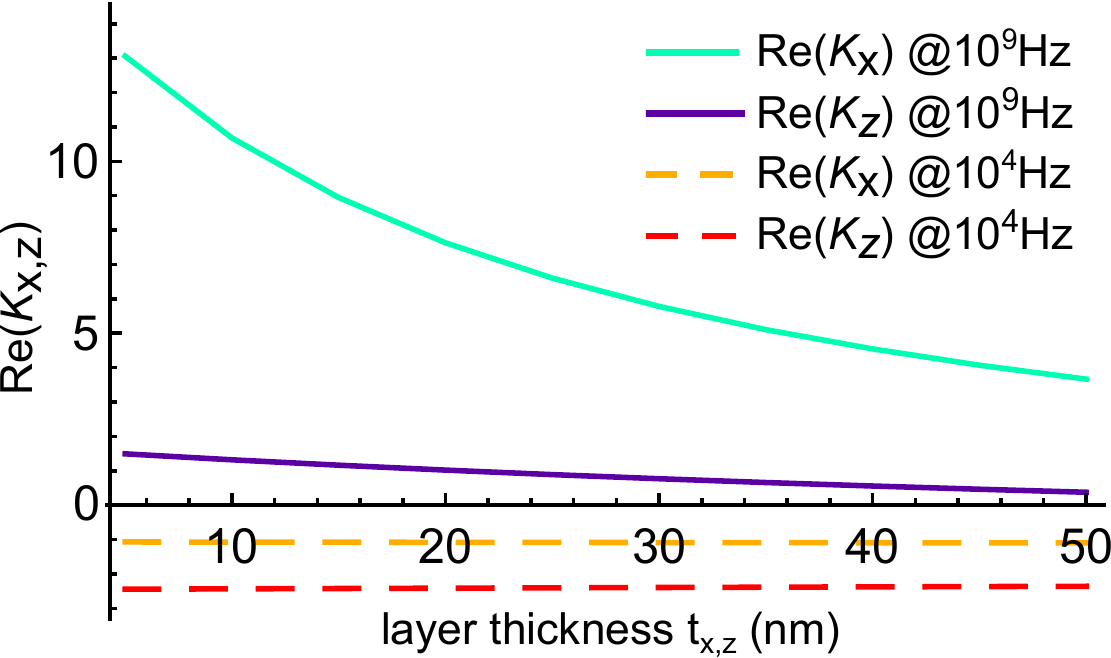}
  \caption{Variation of $\mr{Re}(K_x)$ and $\mr{Re}(K_z)$ with lipid-layer thickness ($t_x$ and $t_z$ respectively) for coated HOPG ellipsoids with $a_2=3\,\mu\mr{m}$, $b_2=0.83\,\mu\mr{m}$ and $c_2=0.5\,\mu\mr{m}$. \textbf{Solid lines}: high-frequency values (at $10^9\,\mr{Hz}$). \textbf{Dashed lines}: low-frequency values (at $10^4\,\mr{Hz}$).}
  \label{fig:KvsLayerThickness}
\end{figure}

In the LCE model, $\mr{Re}(K_x)$ and $\mr{Re}(K_z)$, and therefore the torque, depend significantly on the thickness of the shell along each corresponding direction ($t_x$ and $t_z$ respectively). This is shown in Fig. \ref{fig:KvsLayerThickness} for micro-particle sizes similar to those in our experiments. The high-frequency values of $\mr{Re}(K_{x,z})$ vary substantially with shell thickness, while the low-frequency ones are mostly independent of the shell thickness for the range of parameters explored.

As mentioned earlier, the LCE model uses a confocal ellipsoidal layer with a thickness that varies as a function of position on the ellipsoid surface. The difference with a uniform thickness layer can be significant, as the modeled layer is thinnest along the largest particle dimension and thickest along the smallest particle dimension (Fig. \ref{fig:layeredEllipsoid}). In order to improve the model for uniformly coated particles while maintaining the analytical convenience of the LCE method, we further analyse the dependence of $K_{x,z}$ on the layer thickness. It might be attractive to consider the very-thin-layer approximation (VTLA) as used in earlier studies \cite{T3,AsamiEllipsoidShell,T9}. This stringent approximation is valid for very thin shells with $\delta \ll c_2^2$, where $a_2 \geq b_2 \geq c_2$ and $\delta \approx 2t_x a_2 \approx 2t_y b_2 \approx 2t_z c_2$, leading to the result that $K_k$ depends only on $t_k/p_{1k}$, where $t_k=t_x,t_y,t_z$ and $p_{1k}=a_1,\,b_1,\,c_1$ for $k=x,\,y,\,z$ (see \ref{App:VTLA}). Therefore, in the VTLA, $K_k$ depends only on the ratio of shell thickness to particle size along the $k$ direction. When the VTLA applies, the error introduced by the non-uniform layer thickness is usually considered to be small \cite{T3}. However, this approximation is not valid for our experimental parameters.

In order to model the torque of a uniformly coated particle, it is important to input the correct layer thickness along each direction for each component $K_k$. Gimsa \textit{et al.} \cite{T12,T8} developed a complex finite-element RC model for uniformly coated particles and compared their results to those from the LCE model. They concluded that the LCE model correctly predicts $\mr{Re}(K_k)$ as long as the correct layer thickness is used along the direction of the electric field for thin insulating layers. They tested oblate and prolate ellipsoids with aspect ratios up to 10, with $a_2=b_2=5\,\mu\mr{m}$ and $c_2=0.5\,\mu\mr{m}$ or $c_2=50\,\mu\mr{m}$, and a layer thickness along the field direction equal to $t_x=8\,\mr{nm}$. The same authors suggested the possibility of correcting for the shell-thickness variation of the LCE model by calculating two ellipsoidal models, one for each field component, ensuring that the correct shell thickness is used along each direction \cite{T6}.

As we show in \ref{App:myApprox}, for layers that are thin compared to the particle size, i.e., for the condition $\nu\approx1$, corresponding to $t_x\ll a_2$, $t_y\ll b_2$, $t_z\ll c_2$, it turns out that the equivalent permittivity $\epsilon'_{2k}$ and the effective polarisation factors $K_k$ of the layered ellipsoid along direction $k$ depend mainly on the layer thickness along $k$ and are approximately independent of the layer thickness along the other directions. This enables the convenient separation of the dependencies on the layer thicknesses along the direction of each field component. This separation is very useful for layers that are thin but not quite as thin as for the VTLA to apply, as is the case for our experiments ($\nu \geq 0.8$). For the maximum torque in Eqn.(\ref{eqn:Tmax}), proportional to $\mr{Re}\left(K_x-K_z\right)$, we have that $\mr{Re}(K_x)$ depends mainly on $t_x$ and $\mr{Re}(K_z)$ depends mainly on $t_z$. The choice of the $\delta$ parameter determines the layer thicknesses $t_x$, $t_y$ and $t_z$ for a given ellipsoid. Since $\delta=t_x^2+2t_x a_2=t_y^2+2t_y b_2=t_z^2+2t_z c_2=\mr{const}$ (\ref{App:myApprox}), $\delta$ can be chosen so that the desired layer thickness, $t_\mr{shell}$, is obtained along a chosen direction $k$. For instance, for a desired $t_\mr{shell}$ along $x$, we can choose $\delta=t_\mr{shell}^2+2t_\mr{shell} a_2$ so that $t_x=t_\mr{shell}$ and $t_y,t_z\neq t_\mr{shell}$. Hence, in order to obtain the torque, separate calculations can be carried out for $\mr{Re}(K_x)$ and $\mr{Re}(K_z)$ using two different $\delta$ values ($\delta_x=t_\mr{shell}^2+2t_\mr{shell} a_2$ and $\delta_z=t_\mr{shell}^2+2t_\mr{shell} c_2$, respectively) that each correspond to the desired layer thickness along the $x$ and $z$ directions, respectively.

\subsection{\textbf{One-shell \emph{dual-ellipsoid} model}} \label{sec:2ellipsoidModel}

We propose a new \emph{dual-ellipsoid} model as a good approximation to calculate the torque for an ellipsoid coated with a thin, uniform-thickness shell. This dual-ellipsoid method accounts for the correct shell thickness along each relevant direction for the torque and is valid for ellipsoids with thin shells ($\nu\approx1$). The maximum electro-orientation torque, proportional to $[\mr{Re}(K_x)-\mr{Re}(K_z)]$ [Eqn. (\ref{eqn:Tmax})] is calculated using two different confocal layered ellipsoids. For a desired uniform shell thickness $t_\mr{shell}$, we calculate $\mr{Re}(K_x)$ (that depends mostly on $t_x$) using an ellipsoid with $t_x=t_\mr{shell}$ and $\mr{Re}(K_z)$ (that depends mostly on $t_z$) using an ellipsoid with $t_z=t_\mr{shell}$. In section \ref{sec:results:compareToExperim}, we compare the results of this model to our experimental data. 
\begin{figure}[ht]
	\centering
  \includegraphics[width=0.85\columnwidth]{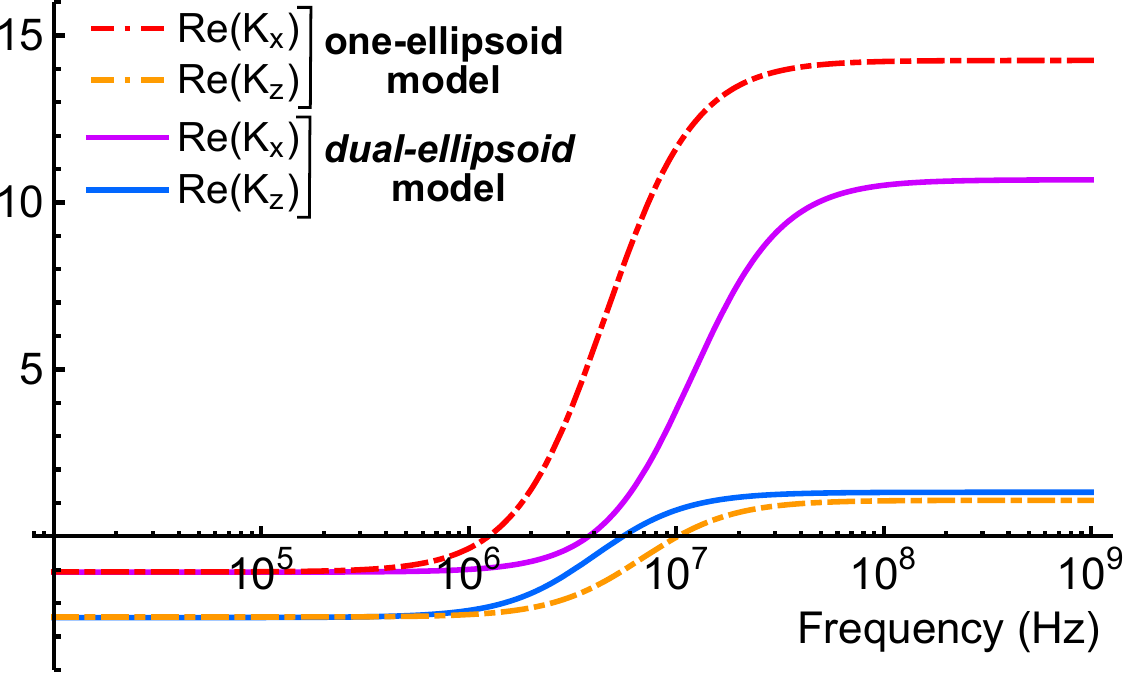}
  \caption{Calculated $\mr{Re}(K_x)$ and $\mr{Re}(K_z)$ versus frequency for a lipid-coated ellipsoidal HOPG particle with $a_2=3\,\mu\mr{m}$, $b_2=0.83\,\mu\mr{m}$ and $c_2=0.5\,\mu\mr{m}$. \textbf{One-ellipsoid model:} same as in Fig. \ref{fig:KxKzVsFreq}, i.e., $t_x\approx3\,\mr{nm}$ and $t_z\approx18\,\mr{nm}$, so $\sim10\,\mr{nm}$ on average. \textbf{\emph{Dual-ellipsoid} model:} $\mr{Re}(K_x)$ is calculated for an ellipsoid with $t_x=10\,\mr{nm}$ and $\mr{Re}(K_z)$ is calculated for a different ellipsoid with $t_z=10\,\mr{nm}$.}
  \label{fig:KxKzVsFreq2}
\end{figure}

\begin{figure}[t]
	\centering
  \includegraphics[width=0.65\columnwidth]{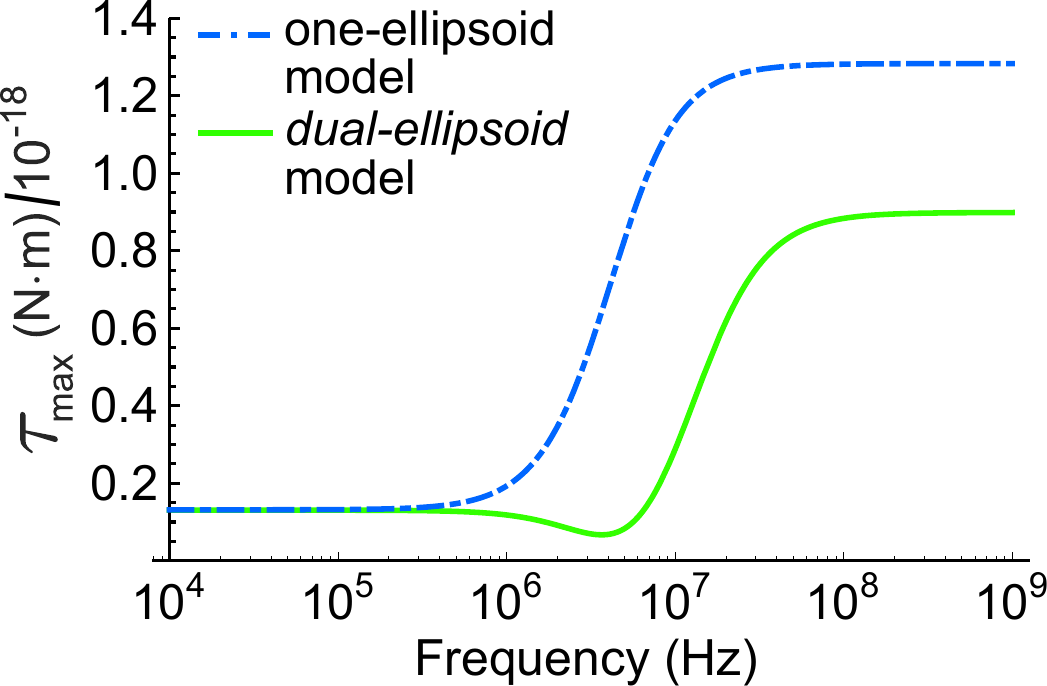}
  \caption{Calculated maximum torque versus frequency for a lipid-coated HOPG ellipsoid with $a_2=3\,\mu\mr{m}$, $b_2=0.83\,\mu\mr{m}$ and $c_2=0.5\,\mu\mr{m}$ and for $t_\mr{shell}=10\,\mr{nm}$. \textbf{One-ellipsoid model:} $\mr{Re}(K_x)$ and $\mr{Re}(K_z)$ both calculated for an ellipsoid with $\delta=(0.14\,\mu \mr{m})^2$ so that $t_x\approx3\,\mr{nm}$ and $t_z\approx18\,\mr{nm}$, so $\sim10\,\mr{nm}$ on average. \textbf{\emph{Dual-ellipsoid} model:} one ellipsoid with $t_x=t_\mr{shell}$ to calculate $\mr{Re}(K_x)$, and a different ellipsoid with $t_z=t_\mr{shell}$ to calculate $\mr{Re}(K_z)$.}
  \label{fig:2models}
\end{figure}

There is a significant difference between using a \emph{dual-ellipsoid} model and using the standard one-ellipsoid model with a layer thickness that is only correct on average, as shown in Figs. \ref{fig:KxKzVsFreq2} and \ref{fig:2models}. For a desired layer thickness $t_\mr{shell}=10\,\mr{nm}$, the one-ellipsoid model uses $\delta=(0.14\,\mu \mr{m})^2$, which corresponds to $t_x\approx3\,\mr{nm}$ and $t_z\approx18\,\mr{nm}$. One could naively think that this is a good model for an average layer thickness value $\sim10\,\mr{nm}$, however, this is not the case, as Fig. \ref{fig:KxKzVsFreq2} shows for $\mr{Re}(K_x)$ and $\mr{Re}(K_z)$. Particularly, the high-frequency values and the positions of the dispersive features are significantly different when comparing the two approaches. This consequently results in substantial differences in the corresponding calculated torque, as shown in Fig. \ref{fig:2models} for the two types of models. The one-ellipsoid model overestimates the torque substantially (except at low frequencies). This is because the thicknesses used, $t_x<t_\mr{shell}$ and $t_z>t_\mr{shell}$, result in an overestimated $\mr{Re}(K_x)$ and an underestimated $\mr{Re}(K_z)$. The dip in the torque in Fig. \ref{fig:2models} in the \emph{dual-ellipsoid} model around $1-10\,\mr{MHz}$ results from the close values of $\mr{Re}(K_x)$ and $\mr{Re}(K_z)$ at the calculated thicknesses in this frequency range (Fig. \ref{fig:KxKzVsFreq2}). The above ideas highlight the need to use the correct layer thickness along each relevant  direction, with the one-shell \emph{dual-ellipsoid} model being closer to the reality of a uniform-thickness layer.

\subsection{\textbf{Theory for ellipsoids with two or more shells}} \label{sec:TwoLayers}

For an ellipsoidal particle coated with two or more shells, we can make use of a nested application \cite{KakutaniEllipsoidShell} of the one-shell LCE theory previously described. The method for two-shell coatings is described in \ref{App:theoryTwoShell} and used to model the behaviour of lipid-coated graphitic micro-flakes with an intermediate aqueous shell between the lipid shell and the HOPG core. Three-shell models can be built in a similar way. Multi-shell ellipsoid models can be used to predict the behaviour of HOPG particles with novel multi-layer coatings, e.g., combinations of lipid layers with layers of other biological molecules (nucleic acids, proteins, specific functionalizations), or of polymers with specific electrical properties. For instance, the conjugation of polymers (e.g., PEG brushes) to the head groups of phospholipids has been used to modify the overall electrical properties of lipid coatings \cite{electrAddressableVesicles}. 
%the adsorption of lipophilic ions to lipid layers has been shown to increase the electric polarizability and capacitance of lipid membranes \cite{T18,T13} 

\section{\label{sec:results} Results from the one-shell \emph{dual-ellipsoid} model}

We now calculate the maximum electric torque [$\mathcal{T}_\mr{max}$ from Eqn.(\ref{eqn:Tmax})] derived from the one-shell \emph{dual-ellipsoid} model as a function of the electric field frequency $f$. Results apply to HOPG micro-flakes coated with a thin lipid shell that have their planes vertically pre-aligned and that are subjected to a horizontal AC electric field. We highlight the parameters that most critically influence the frequency dependency of the torque acting on the particles, i.e., the thickness of the lipid shell, the conductivity of the solution and the shape of the particle. Finally, we compare our results to our experimental data for lipid-coated HOPG micro-particles in 20 mM NaCl aqueous solution \cite{magnetoElectrOrientHOPG}. Results from the two-shell \emph{dual-ellipsoid} model are briefly shown in \ref{App:theoryTwoShell}.

\subsection{\label{sec:results:changeLayerThickness} \textbf{Varying the thickness of the lipid shell}}

The one-shell \emph{dual-ellipsoid} model calculations assume ellipsoidal HOPG micro-particles with sizes similar to those in our experiments (see Table \ref{tab:simulParams}). The electrical properties used for the 20 mM NaCl aqueous solution are those expected at $25^{\circ}\mr{C}$ \cite{NaClsolutionProps}: $\varepsilon_1=80$ and $\sigma_1=0.2\,\mr{S/m}$. For the lipid shell, we use the permittivity and conductivity values reported for lipid-bilayer membranes in \cite{TBJonesBook}: $\varepsilon_2=11$ and $\sigma_2=10^{-7}\,$S/m. For the dielectric properties of HOPG we use the optical functions reported by Jellison \textsl{et al.} \cite{HOPGdielectricProperties}, obtained from visible light ellipsometry measurements: we input the average values $\varepsilon_{\mr{hopg},\parallel}=2.4$, $\sigma_{\mr{hopg},\parallel}=2.5 \times 10^6\,$S/m, $\varepsilon_{\mr{hopg},\perp}=1.75$ and $\sigma_{\mr{hopg},\perp}=250\,$S/m. We input an electric field amplitude $E_0 = 10^4\,\mr{V/m}$.
\begin{figure}[ht]
	\centering
  \includegraphics[width=0.9\columnwidth]{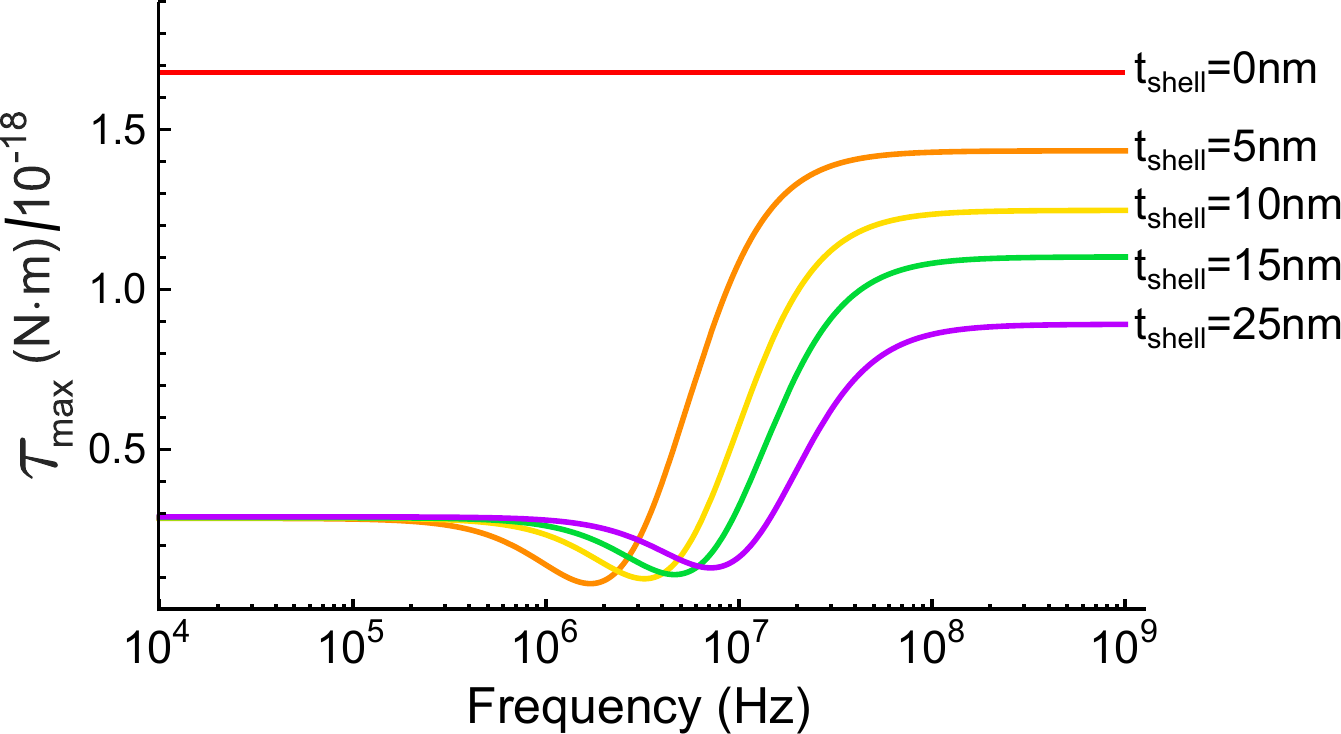}
  \caption{Maximum torque versus electric-field frequency calculated with the one-shell \emph{dual-ellipsoid} model for a lipid-coated HOPG micro-flake ($a_2=3\,\mu \mathrm{m}$, $b_2=1.3\,\mu \mathrm{m}$, $c_2=0.6\,\mu \mathrm{m}$) for different lipid-shell thicknesses, $t_\mr{shell}$: $0\,\mr{nm}$, $5\,\mr{nm}$, $10\,\mr{nm}$, $15\,\mr{nm}$ and $25\,\mr{nm}$.}
  \label{fig:calcTvsFreq_LipidLayer}
\end{figure}

Figure \ref{fig:calcTvsFreq_LipidLayer} shows the calculated maximum torque on a particle as a function of frequency for various thicknesses of the lipid shell from $0\,\mr{nm}$ to $25\,\mr{nm}$. As previously noted, the electrically insulating lipid coating reduces the electric torque acting on the particles, particularly at frequencies below $\sim10^6\,\mr{Hz}$. Increasing the thickness of the lipid shell results in a decrease in shell capacitance, so that larger frequencies are required for shell bridging, shifting the main dispersion feature (low-to-high torque transition region) to higher frequencies. A smoother dispersion is observed for increasing shell thicknesses owing to the marked decrease in high-frequency torque for thicker shells.    

\subsection{\label{sec:results:changeSolutionConductivity} \textbf{Dependency on the  solution conductivity}}

\begin{figure}[t]
	\centering
  \includegraphics[width=0.76\columnwidth]{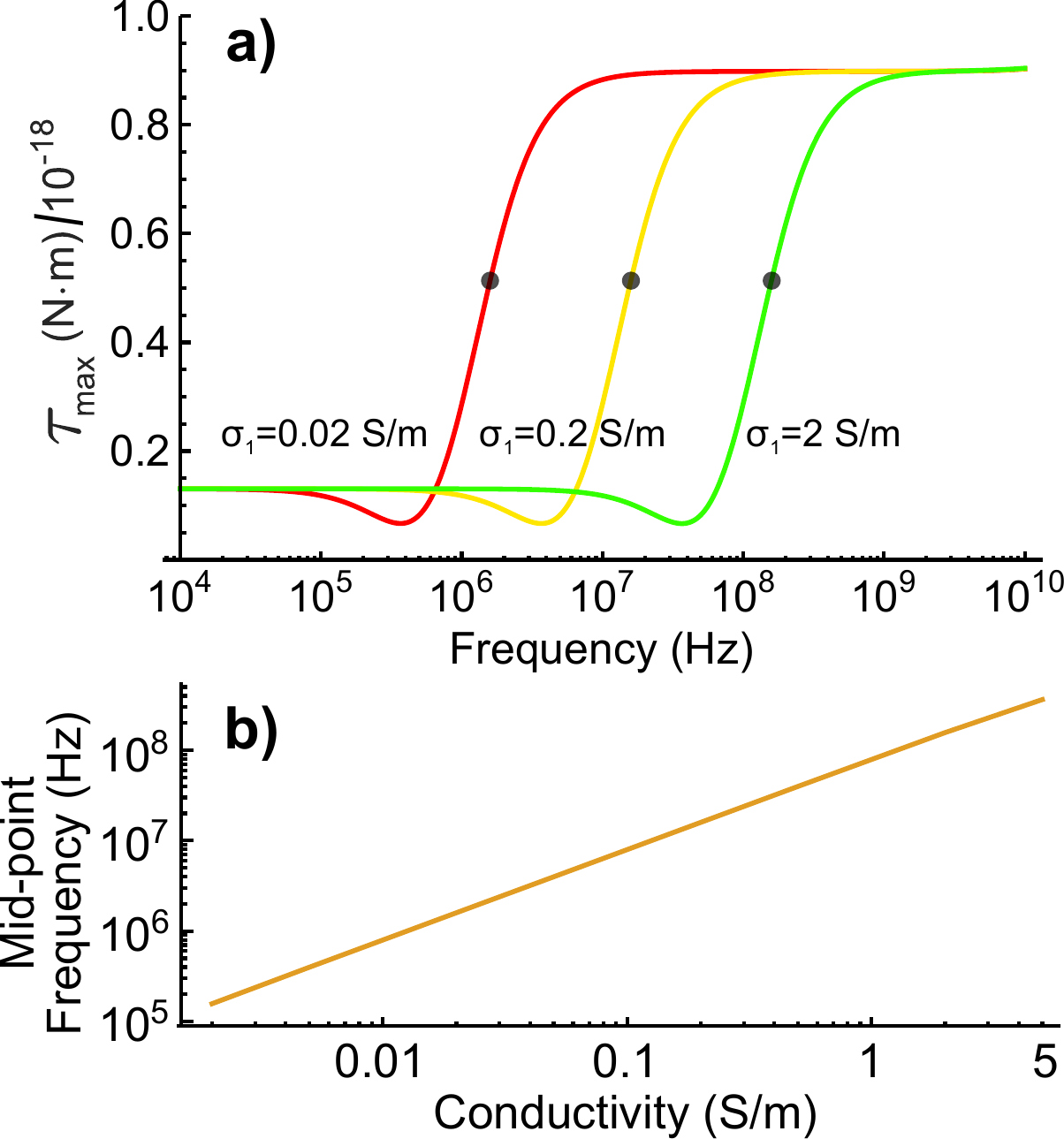}
  \caption{\textbf{a)} Maximum electric torque versus frequency for a lipid-coated HOPG particle ($a_2=3\,\mu \mathrm{m}$, $b_2=0.83\,\mu \mathrm{m}$, $c_2=0.5\,\mu \mathrm{m}$) for different conductivities $\sigma_1$ of the surrounding fluid: 2 S/m, 0.2 S/m and 0.02 S/m. For all curves, $t_\mr{shell}=10\,\mr{nm}$. \textbf{b)} Corresponding mid-point frequency (filled circles in a)) versus solution conductivity.}
  \label{fig:calcTvsFreq_FluidConductivity}
\end{figure}

The torque exerted on the lipid-coated HOPG micro-particle is extremely sensitive to the electrical conductivity of the solution. Figure \ref{fig:calcTvsFreq_FluidConductivity}a shows the predicted torque for solution conductivities $0.02\,\mr{S/m}$, $0.2\,\mr{S/m}$ and $2\,\mr{S/m}$, corresponding to 2 mM, 20 mM and 200 mM concentrations of NaCl in water, respectively \cite{NaClsolutionProps} (the remaining parameters are the same as in section \ref{sec:results:changeLayerThickness}). The main dispersion feature moves to higher frequencies with increasing solution conductivity $\sigma_1$. As the main dispersion feature is due to the frequency-dependent capacitive impedance of the lipid shell, when $\sigma_1$ increases, the difference in conductivity between the solution and the lipid shell increases and the larger accumulation of free charges in the electrolyte solution on the outer side of the lipid shell has a shielding effect diminishing the effective lipid-shell capacitance and requiring higher frequencies for capacitive bridging of the shell. This effect explains the shift to higher frequencies of the mid-frequency-point in the dispersion for increasing solution conductivities, as shown in Fig. \ref{fig:calcTvsFreq_FluidConductivity}b. This property offers a useful control handle for electro-orientation experiments. Furthermore, given the high sensitivity of these layered micro-particles to the conductivity of their environment, one could think of using them as probes of local ionic strength, for instance, near cell surfaces, or in chemical sensing applications.

\subsection{\label{sec:results:changeParticleSize} \textbf{Dependency on particle shape}}

Modifying particle shape and geometry can also have a substantial effect. Since the torque depends linearly on the particle volume (Eqn. \ref{eqn:Tmax}), we examine the effect of changing shape at constant volume. The parameters used are those given at the start of section \ref{sec:results:changeLayerThickness} and a lipid-shell thickness $t_\mr{shell}=10\,\mr{nm}$ is used for all cases.

\begin{figure}[ht]
	\centering
  \includegraphics[width=0.67\columnwidth]{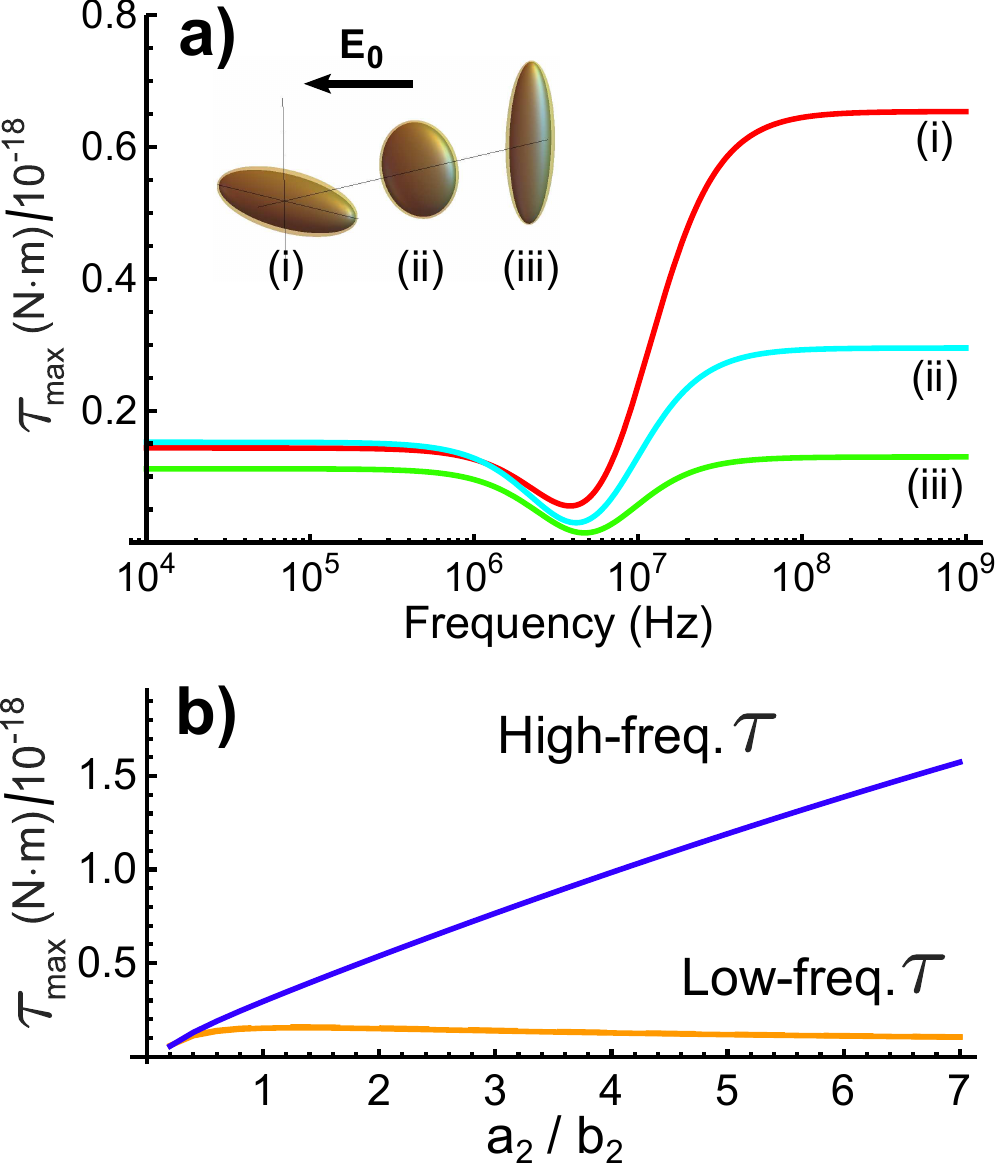}
  \caption{\textbf{a)} $\mathcal{T}_\mr{max}$ versus frequency calculated with the one-shell \emph{dual-ellipsoid} model for different ratios $a_2/b_2$ (particle eccentricity on the $x$-$y$ plane), for ellipsoids of equal volume: (i) $a_2=2.5\,\mu\mr{m}$, $b_2=1\,\mu\mr{m}$ and $c_2=0.5\,\,\mu\mr{m}$, (ii) $a_2=b_2=1.58\,\mu\mr{m}$ and $c_2=0.5\,\,\mu\mr{m}$ and (iii) $a_2=1\,\mu\mr{m}$, $b_2=2.5\,\mu\mr{m}$ and $c_2=0.5\,\,\mu\mr{m}$. The layer thickness is set to $t_\mr{shell}=10\,\mr{nm}$. \textbf{b)} High-frequency (at $f=10^9$ Hz) and low-frequency (at $f=10^4$ Hz) torque as a function of $a_2/b_2$.}
  \label{fig:calcTvsFreq_Eccentricity}
\end{figure}

Figure \ref{fig:calcTvsFreq_Eccentricity} shows the effect of changing the eccentricity of the particle on the $x$-$y$ plane. Curves corresponding to three different ellipsoidal geometries with equal volume, equal size along $z$ ($c_2=0.5\,\,\mu\mr{m}$) and different eccentricity ratio $a_2/b_2$ are shown: (i) $a_2=2.5\,\mu\mr{m}$ and $b_2=1\,\mu\mr{m}$, (ii) $a_2=b_2=1.58\,\mu\mr{m}$ and (iii) $a_2=1\,\mu\mr{m}$ and $b_2=2.5\,\mu\mr{m}$. 
Figure \ref{fig:calcTvsFreq_Eccentricity}a shows that $\mathcal{T}_\mr{max}$ at high frequencies increases when the particle size increases along the in-plane $x$ direction. Figure \ref{fig:calcTvsFreq_Eccentricity}b shows this high-frequency torque increasing with an increasing ratio $a_2/b_2$. The torque at low frequencies, on the other hand, shows very little variation with $a_2/b_2$. The fact that the lipid-shell thickness relative to the particle size along $x$ (i.e., the ratio $t_x / a_2$) is increasing from curve (i) to curve (iii) in Fig. \ref{fig:calcTvsFreq_Eccentricity}a, also contributes to the tendency shown. However, this contribution is small and the observed tendency remains when modifying the lipid thickness to keep a constant $t_x / a_2$ ratio for all particle geometries (see \ref{App:changeEccentrSuppl}).

\begin{figure}[t]
	\centering
  \includegraphics[width=0.72\columnwidth]{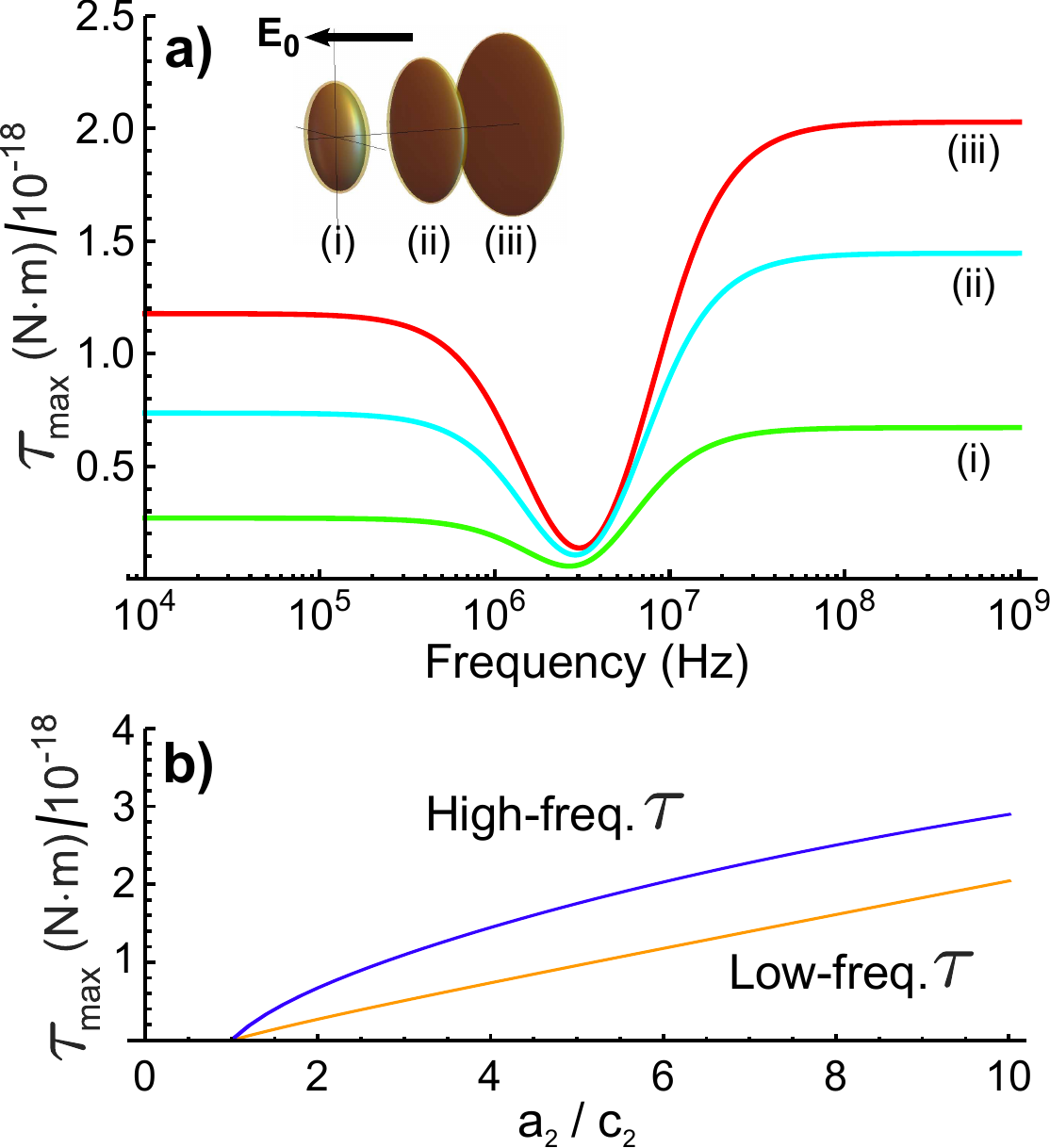}
  \caption{\textbf{a)} $\mathcal{T}_\mr{max}$ versus frequency calculated with the one-shell \emph{dual-ellipsoid} model for different particle aspect ratios, $a_2/c_2$, for oblate ellipsoids ($a_2=b_2$) of equal volume: (i) $a_2=b_2=3\,\mu\mr{m}$, $c_2=0.5\,\,\mu\mr{m}$, (ii) $a_2=b_2=2.62\,\mu\mr{m}$, $c_2=0.66\,\,\mu\mr{m}$ and (iii) $a_2=b_2=2.08\,\mu\mr{m}$, $c_2=1.04\,\,\mu\mr{m}$. The layer thickness is set to $t_\mr{shell}=10\,\mr{nm}$. \textbf{b)} Low-frequency torque (at $f=10^4$ Hz) and high-frequency (at $f=10^9$ Hz) torque versus $a_2/c_2$.}
  \label{fig:calcTvsFreq_AspectRatio}
\end{figure}

The effect of modifying the particle aspect ratio $a_2/c_2$ for oblate ellipsoids ($a_2=b_2$) is shown in Fig. \ref{fig:calcTvsFreq_AspectRatio}. In Fig. \ref{fig:calcTvsFreq_AspectRatio}a, we plot $\mathcal{T}_\mr{max}$ for three particles of equal volume and different aspect ratios $a_2/c_2 = 6$, $a_2/c_2 = 4$ and $a_2/c_2 = 2$, corresponding respectively to: (i) $a_2=b_2=3\,\mu\mr{m}$, $c_2=0.5\,\,\mu\mr{m}$; (ii) $a_2=b_2=2.62\,\mu\mr{m}$, $c_2=0.66\,\,\mu\mr{m}$; (iii) $a_2=b_2=2.08\,\mu\mr{m}$, $c_2=1.04\,\,\mu\mr{m}$. Fig. \ref{fig:calcTvsFreq_AspectRatio}b shows the high-frequency and low-frequency torque as a function of $a_2/c_2$. Increasing the particle aspect ratio at constant volume for oblate geometries leads to an increased torque at all frequencies. This is due to the fact that, at constant layer thickness, decreasing $a_2/c_2$ increases the relative layer thickness ($t_x/a_1$) along $x$ and decreases it ($t_z/a_1$) along $z$, hence reducing $\mr{Re}(K_x)$ and increasing $\mr{Re}(K_z)$, respectively. 

Our model also shows that particle-shape anisotropy plays a dominant role in the resulting torque compared to electrical anisotropy, given that results are unmodified when electrically isotropic particles are considered (using either the HOPG in-plane conductivity, $\sigma_3=2.5 \times 10^6\,$S/m, or the out-of-plane value $\sigma_3=250\,$S/m). This is because the difference in conductivity between HOPG (in-plane or out-of-plane) and the solution (and lipid shell) is so large, that the torque is unmodified unless this difference is substantially reduced (see \ref{App:changeHOPGcond}).

\subsection{\label{sec:results:compareToExperim} \textbf{Comparison with experimental data}}

Our previous report \cite{magnetoElectrOrientHOPG} presented measurements of the maximum electric torque acting on 10 different lipid-coated HOPG micro-flakes submerged in 20 mM NaCl aqueous solution. A sample chamber with two parallel thin wire electrodes ($50\,\mu\mr{m}$-diameter copper wires at a centre-to-centre distance of $150\,\mu\mr{m}$) glued onto a glass coverslip (Fig. \ref{fig:EfieldWires}a in \ref{App:EfieldWires}) was used to expose individual HOPG particles to a horizontal AC electric field. Particles between the two wires were vertically pre-aligned using a vertical magnetic field and imaged edge on from above with a custom-built microscope \cite{magnetoElectrOrientHOPG}. Measurements were taken at room temperature at frequencies between 1 and $70\,\mr{MHz}$, with several measurements per frequency value and micro-flake. Particle rotation was observed mostly at frequencies above $\sim10\,\mr{MHz}$ in experiments. When the AC electric field was turned on, a given HOPG micro-flake rotated to align parallel to the electric field direction. Once aligned, the micro-flake was rotationally trapped, i.e., the orientation of its graphene planes was confined to the vertical plane defined by both the vertical magnetic field and horizontal AC electric field. The maximum electric torque, $\mathcal{T}_\mr{max}$, was obtained by analysing the measured orientational Brownian fluctuations around the equilibrium orientation in the rotational traps. A fit of the autocorrelation of such fluctuations yielded the rotational trap stiffness. Given that the particles were in close proximity to the glass substrate, in order to eliminate the effect of interactions with the substrate, the rotational trap stiffness obtained in the absence of an electric field ($k_\mr{off}$) was subtracted from that measured immediately after in the presence of the electric field ($k_\mr{on}$). $\mathcal{T}_\mr{max}$ was then obtained as $(k_\mr{on}-k_\mr{off})/2$ (refer to \cite{magnetoElectrOrientHOPG} for details).

We apply a small correction to the raw torque data in order to account for a small observed variation with frequency of the input voltage after amplification. This ensures that we can compare to simulations that use a constant $E_0$ for all frequencies. Given that the electric torque is proportional to the square of the electric field amplitude (Eqn. \ref{eqn:Tmax}) and therefore to the square of the voltage, the raw torque measurements are multiplied by the correction factor $(V_\mr{in}/V_\mr{in,DC})^{-2}$, where $V_\mr{in}$ is the is input AC voltage amplitude into the electrode wires and  and $V_\mr{in,DC}$ is the input voltage amplitude at zero frequency. Figure \ref{fig:EfieldWires}c in \ref{App:EfieldWires} shows a plot of $V_\mr{in}/V_\mr{in,DC}$ as a function of frequency.

\begin{figure}[t]
	\centering
	\includegraphics[width=0.8\columnwidth]{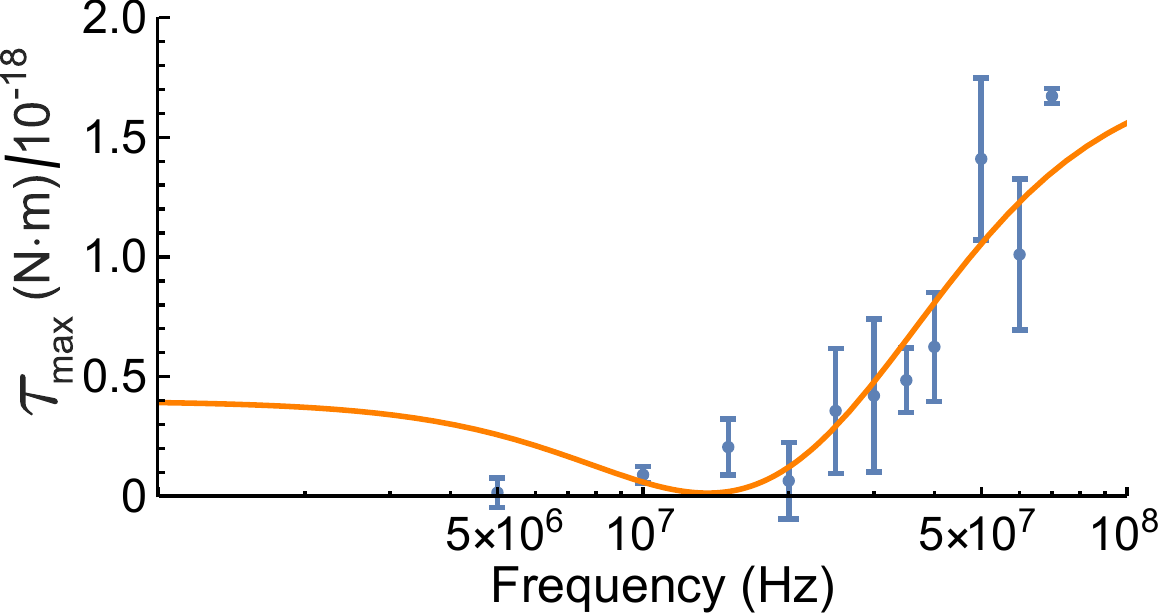}
  \caption{$\mathcal{T}_\mr{max}$ versus frequency for a lipid-coated HOPG micro-flake in 20 mM NaCl aqueous solution: \textbf{particle A} (Table \ref{tab:simulParams}). \textbf{Data points:} experimental measurements \cite{magnetoElectrOrientHOPG}. The data points and error bars correspond to the mean and standard deviation, respectively, for several measurements at each frequency. \textbf{Solid line:} fitted theoretical curve to the one-shell \emph{dual-ellipsoid} model.}
	\label{fig:fitDataParticleA}
\end{figure}

\begin{figure}[ht]
	\centering
	\includegraphics[width=0.8\columnwidth]{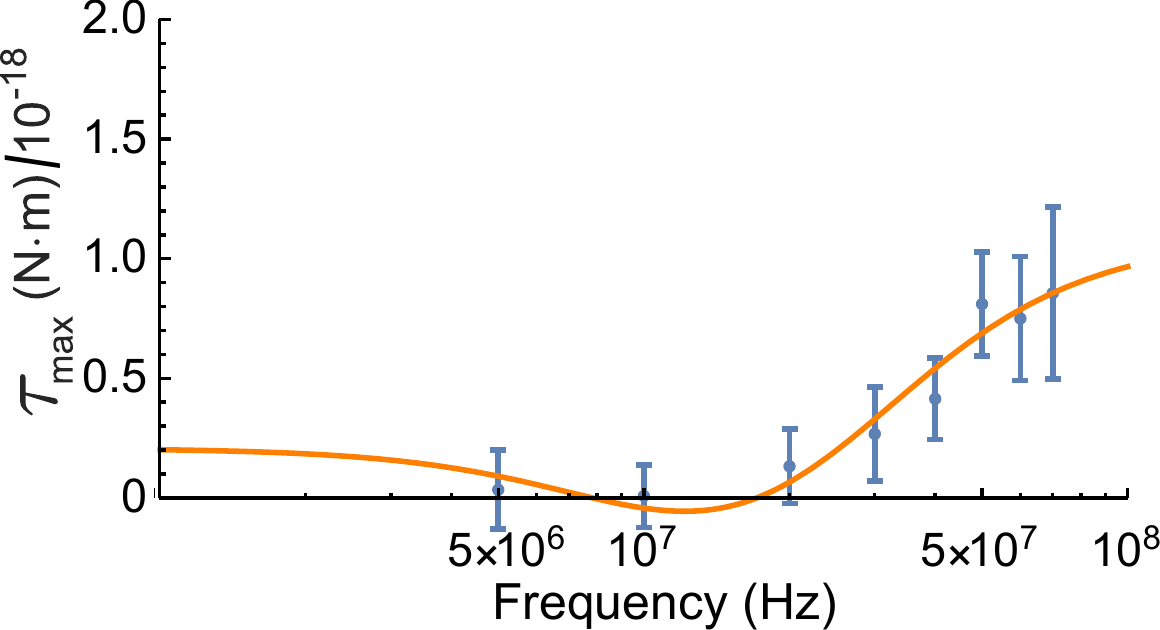}
  \caption{$\mathcal{T}_\mr{max}$ versus frequency for \textbf{particle B} (Table \ref{tab:simulParams}). \textbf{Data points:} experimental measurements. \textbf{Solid line:} fitted theoretical curve to the one-shell \emph{dual-ellipsoid} model.}
	\label{fig:fitDataParticleB}
\end{figure}

\begin{figure}[ht]
	\centering
	\includegraphics[width=0.8\columnwidth]{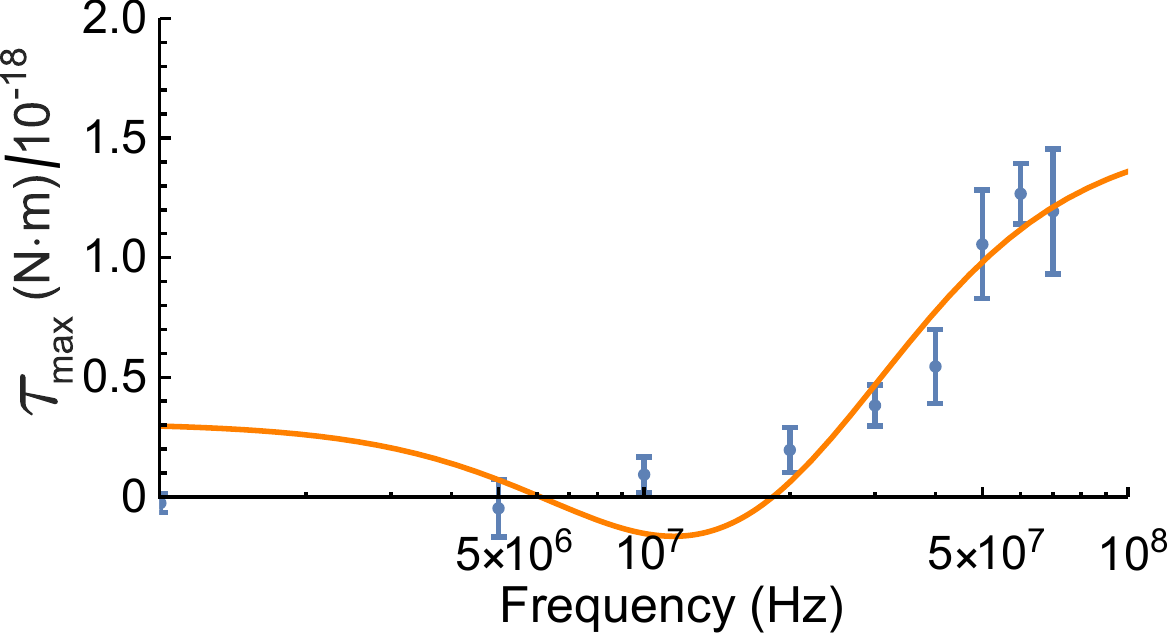}
  \caption{$\mathcal{T}_\mr{max}$ versus frequency for \textbf{particle C} (Table \ref{tab:simulParams}). \textbf{Data points:} experimental measurements. \textbf{Solid line:} fitted theoretical curve to the one-shell \emph{dual-ellipsoid} model.}
	\label{fig:fitDataParticleC}
\end{figure}

\begin{figure}[ht]
	\centering
	\includegraphics[width=0.8\columnwidth]{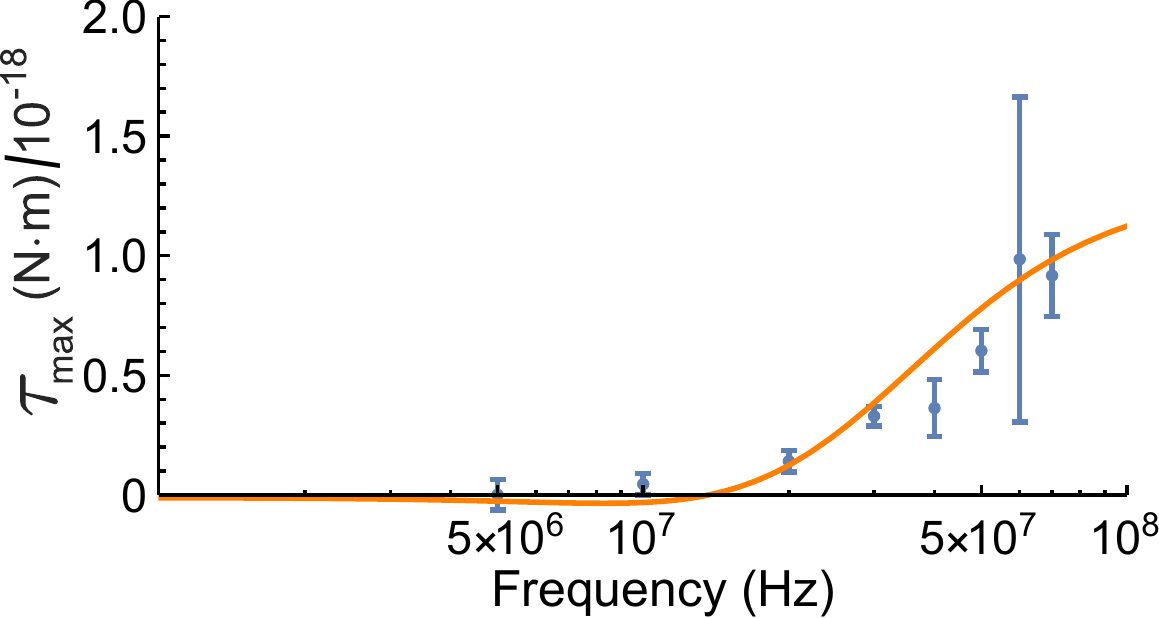}
  \caption{$\mathcal{T}_\mr{max}$ versus frequency for \textbf{particle D} (Table \ref{tab:simulParams}). \textbf{Data points:} experimental measurements. \textbf{Solid line:} fitted theoretical curve to the one-shell \emph{dual-ellipsoid} model.}
	\label{fig:fitDataParticleD}
\end{figure}

For all the particles measured, $\mathcal{T}_\mr{max}$, and therefore the strength of the rotational trapping, increased with increasing electric field frequency above $10\,\mr{MHz}$ \cite{magnetoElectrOrientHOPG}. In order to compare the experimental data with our one-shell \emph{dual-ellipsoid} model, it is best to look at measurements for individual particles as opposed to averages, given the important dependency on particle shape and orientation shown in section \ref{sec:results:changeParticleSize}. The data points in Figs. \ref{fig:fitDataParticleA} to \ref{fig:fitDataParticleD} show the measured maximum electric torque for four particles. These are shown as examples and compared to the model predictions. The error bars in the plots correspond to the standard deviation of the several measurements (typically 5) performed at each frequency. The data for all HOPG micro-particles can be found in \ref{App:allTorqueData}. 

From a first look at Figures \ref{fig:fitDataParticleA} to \ref{fig:fitDataParticleD}, we can see that the presence of an increasing torque with increasing field frequency above $\sim 10\,\mr{MHz}$ is clearly in agreement with the main dispersive feature predicted by our models. However, we can also see that the torque measured at the lowest frequency values, $1-10\,\mr{MHz}$, has a value close to zero for all particles. This is not in full agreement with the predictions from our models for our particle shapes and sizes, from which we would expect a non-zero low-frequency torque. It is a physically intuitive expectation that polarisable and shape-anisotropic particles will align for even DC electric fields. It is therefore likely that the presence of non-trivial interactions of the particle with the nearby glass substrate leads to a lower measured torque than expected. We therefore introduce a constant torque offset ($\mathcal{T}_\mr{max,offset}$) as a fitting parameter in order to compare with our theoretical model. The data points are fitted to the model using as fitting parameters the field amplitude ($E_0$), the static relative permittivity of the lipid shell ($\varepsilon_2$), the lipid-layer thickness ($t_\mr{shell}$) and the offset ($\mathcal{T}_\mr{max,offset}$). A non-linear fit is performed using the error bars (standard deviation) of the data points for weighted fitting (reasonable bounds are applied as parameter constraints and differential evolution is used as minimization method). The remaining relevant parameters are fixed and take values (see \ref{sec:results:changeLayerThickness}): $\varepsilon_1=80$ \cite{NaClsolutionProps}, $\sigma_2=10^{-7}\,$S/m \cite{TBJonesBook}, and $\varepsilon_{\mr{hopg},\parallel}=2.4$, $\sigma_{\mr{hopg},\parallel}=2.5 \times 10^6\,$S/m, $\varepsilon_{\mr{hopg},\perp}=1.75$ and $\sigma_{\mr{hopg},\perp}=250\,$S/m \cite{HOPGdielectricProperties}. A solution conductivity of $\sigma_1 = 0.3\,\mr{S/m}$, slightly higher than the value of $0.2\,\mr{S/m}$ expected for a 20 mM NaCl solution \cite{NaClsolutionProps}, is found to best fit the experimental data. The dimensions of all particles are given in Table \ref{tab:simulParams}. Note that $\varepsilon_2$ is taken as a fitting parameter owing to the fact that various references report different values for it: $\sim 11$ for cell membranes \cite{TBJonesBook}, $\sim 6$ for the lipid bilayer membrane of retinal rods \cite{RetinalRods}, $\sim 3$ for DPPC bilayer patches \cite{DPPCbilayers}, and $\sim 2$ for bilayer membranes \cite{permittBilayer}. 
\begin{table*}[ht]
	\centering
	\begin{tabular}{|c||c|c|c|c|}
		\hline
		Particle label & A (0106\_p2) & B (0906\_p4) & C (1106\_p3) & D (2406\_p1) \\
		\hline
				$a_2$ ($\mu\mr{m}$) & 3.4 & 3.0 & 3.4 & 3.1 \\
		\hline
				$b_2$ ($\mu\mr{m}$) & 1.4 & 1.3 & 1.6 & 0.7 \\
		\hline
				$c_2$ ($\mu\mr{m}$) & 0.6 & 0.6 & 0.6 & 0.7 \\
		\hline
				volume ($\mu\mr{m}^3$) & 13.1 & 10.2 & 13.8 & 6.4 \\
		\hline
				substrate & plasma cleaned & plasma cleaned & PEGylated & untreated \\
		\hline
				$E_0$ ($\mr{V/m}$) & $1.4\times 10^4$ & $1.3\times 10^4$ & $1.2\times 10^4$ & $1.5\times 10^4$ \\
	  \hline
		    $\varepsilon_2$ & 4.5 & 9.1 & 6.3 & 6.6 \\
		\hline
				$t_\mr{shell}$ ($\mr{nm}$) & 15 & 24 & 17 & 16 \\
		\hline
			  $\mathcal{T}_\mr{max,offset}$	($\mr{N}.\mr{m}/10^{-18}$) & -0.4 & -0.3 & -0.5 & -0.2 \\
		\hline
		    R-squared of fit &	0.86 & 0.95 & 0.87 & 0.89   \\
		\hline
	\end{tabular}
	\caption{Properties of four measured lipid-coated HOPG micro-flakes (A, B, C and D) in 20mM NaCl aqueous solution above a glass substrate. $E_0$, $t_\mr{layer}$, $\varepsilon_2$ and $\mathcal{T}_\mr{max,offset}$: values obtained from fits of the data (Figs. \ref{fig:fitDataParticleA} to \ref{fig:fitDataParticleD}) to the one-shell \emph{dual-ellipsoid} model.}
	\label{tab:simulParams}
\end{table*}

The values obtained from the fits for $E_0$, $\delta$ and $\mathcal{T}_\mr{max,offset}$ for each of the particles are shown in Table \ref{tab:simulParams}. The fitted $E_0$ takes values between $1.2\times 10^4\,\mr{V/m}$ and $1.5\times 10^4\,\mr{V/m}$. These are reasonably close to the value $E_0 \sim 1.5 \times 10^4\,\mr{V/m}$ expected at the sample region in between the electrode wires assuming perfect impedance matching (see field-map calculation in \ref{App:EfieldWires}). Impedance mismatch effects in the electrical connections would reduce the field value at the sample compared to that value. Determination of the exact AC MHz electric field amplitude at the sample is non-trivial, this being the main reason for making $E_0$ a fitting parameter. The obtained fitted lipid-layer thickness values, $t_\mr{shell}$, range from $15\,\mr{nm}$ to $24\,\mr{nm}$ (see Table \ref{tab:simulParams}). These correspond roughly to 3-5 lipid bilayers coating each particle ($\sim 5\,\mr{nm}$ per bilayer). This agrees very well with our previous atomic force microscopy images and indentation measurements that showed 4-5 indentation steps corresponding to several lipid bilayers coating a HOPG micro-particle \cite{magnetoElectrOrientHOPG}. The R-squared values for the fits are between 0.86 and 0.95 (Table \ref{tab:simulParams}). Overall, our one-shell \emph{dual-ellipsoid} model with the addition of an offset reproduces well the increase in torque with frequency measured for submerged HOPG micro-flakes in the range 10-70 MHz and yields realistic values for the fitted parameters that agree reasonably well with the experimental ones.\\

The need for the offsets in the fits shows that it is likely that the interaction of the lipid-coated HOPG micro-particles with the nearby glass substrate is more complex than initially thought, leading to a reduction of the measured torque on the micro-flakes. In order to analyse the data, we modelled the glass-particle interaction as an additional rotational trapping effect. This was motivated by the fact that the mean squared displacement of the measured orientational fluctuations at zero field was non-linear, pointing to the presence of weak trapping due to the presence of the substrate, even in the absence of the field. More complex interaction effects (e.g., a combination of trapping, increased friction, sticking and/or perturbation of electric field lines) would not be easily decoupled from the effect of the electric field in measurements of particle orientational fluctuations. Glass surfaces in liquid can typically present surface charges that could locally modify the direction of the electric field lines, possibly reducing the horizontal field component that aligns the micro-flakes between the electrode wires and lowering the measured torque values. Mobile glass-surface charges could also in principle generate complex frequency-dependent effects at the glass-particle interface, which would be beyond the scope of our model. As shown in \ref{App:theoryTwoShell}, the presence of an intermediate aqueous solution layer between the HOPG core and the thin lipid shell would not explain the measured zero low-frequency torque.

\section{\label{sec:Conclusions} Discussion and conclusions}

We have presented a detailed theoretical study of the electro-orientation of lipid-coated graphitic micro-particles in aqueous solution in the presence of an AC electric field. The use of graphitic micro-particles dispersed in solution can have numerous applications in the fields of materials science, opto-electronics, energy storage, biophysics and chemical and biological sensing. These particles are dispersed in solution typically via the use of surfactant layers or, as in this work, coating the micro-particles with a lipid shell. The overall electrical properties of the layered particles in solution can be substantially different to those of the uncoated particles. The models presented in this work can be used to predict the (reduced) efficiency of the electrical manipulation of the layered micro-particles and the dependency of this effect on the frequency of the applied electric field. In this way, the particle shapes and coatings, frequency ranges and experimental conditions at which electro-orientation can be observed can be predicted for successful experimentation. In particular, the work presented in this paper sets the ground for the use of lipid-coated HOPG micro-flakes in aqueous solution as a tool to quantitatively sense biologically relevant torque and rotary motion in biophysical studies. These layered graphitic particles are biocompatible, can be used in biocompatible solutions, can be functionalised and attached to relevant bio-molecules, and are therefore ideal for such application. 

We have presented a new \emph{dual-ellipsoid} model based on the Laplace confocal ellipsoid theory and accounting for the correct layer thickness along each direction relevant for the torque calculation. The model is valid for thin layers and moderate aspect ratios. Our theory is more detailed than models found in previous reports for the electro-orientation of graphite/graphene micro-flakes \cite{grapheneGraphiteFlakeElectricAlign2013,grapheneOxideMonolayersElectrAlign2014}, that did not consider layered particles. Modelling the particles as layered ellipsoids, we have calculated the electric torque acting on them when they are exposed to MHz AC electric fields in solution. Our models consider the electric conductivity and permittivity of the HOPG core, the lipid shell and the solution, and account for Maxwell-Wagner interfacial polarisation effects at all interfaces, successfully predicting the dependency with frequency of the electric torque. We have identified that the torque depends crucially on the thickness of the lipid shell, on the conductivity of the solution and on the shape of the particle. These provide useful handles to control the outcome of experiments. 

Our theoretical predictions are in good quantitative agreement with the experimental data for HOPG micro-flakes in 20mM NaCl aqueous solution. Results highlight the fact that electro-orientation may be observed experimentally only at certain electric field frequencies. The predictions from our model with the addition of an offset reproduce well the measured increase in torque in the frequency range 10-70 MHz. Parameter values yielded by fits of the data to the model are realistic and agree well with the experimental ones. Given the lack of quantitative torque predictions for electro-orientation in the literature, this is an important advancement.

\section*{Acknowledgments}
The authors would like to thank Prof. Sonia Contera for extremely useful discussions. We are indebted to HEFCE and EPSRC for their funding support.

%\begin{acknowledgments}
%\end{acknowledgments}

\appendix

\section{Dipole-dipole interactions and particle concentration} \label{App:particleDistance}

The potential energy of interaction between two aligned, stationary electric dipoles is attractive, proportional to the product of the dipole moments and short-ranged, decaying as $1/r^3$, where $r$ is the inter-particle (centre-to-centre) distance \cite{PhysChemLifeBook}. It follows that the dipole-dipole force decays as $1/r^4$. The attraction between aligned induced dipoles as well as the local distortion of the external applied AC field in the vicinity of the particles can lead to the formation of particle chains \cite{TBJonesBook} or ordered colloidal crystals \cite{ACbeadCrystal}. Considering spherical particles of radius $R$ in solution, interactions become very strong for very close distances $r<2.1R$ when the ratio of the permittivity or conductivity of the particles to that of the solution is above $\sim 10$ \cite{TBJonesBook}. When particle chains are formed, the effective dipole moment per particle is greatly enhanced by the interactions and the enhancement factor increases with the number of particles in the chain. As the dipole-dipole interactions decay fast with $r$, they can be considered negligible at inter-particle distances $r\geq 4R$. For example, this has been recently demonstrated by optical-tweezer measurements of induced dipole-dipole forces between polystyrene micro-spheres in aqueous solution in AC electric fields \cite{dipoleDipoleInteractionsBeads}. 

Considering a uniform dispersion of particles, the volume fraction ($\phi = V_{\mr{particles}}/V_{\mr{solution}}$) or weight fraction ($w = m_{\mr{particles}}/m_{\mr{solution}}$) is typically given to characterise the concentration of particles in solution. For spherical particles (case I), the inter-particle distance, $r$, in the dispersion can be considered to be twice the radius of a sphere surrounding each particle, with that sphere having a volume equal to $v = V_{\mr{solution}}/N=n^{-1}$, i.e., the corresponding volume per particle, where $N$ is the number of particles in the solution and $n$ is the number density. Therefore, $r=2 \left[ 3/(4\pi n) \right]^{1/3}$. The number density can be related to the volume fraction or weight fraction by: $n=\phi/V_{\mr{sphere}}$ and $n=(\rho_{\mr{liquid}}/\rho_{\mr{particle}})\, w /V_{\mr{sphere}}$, where $V_{\mr{sphere}}=(4/3)\pi R^3$ is the volume of each spherical particle and $\rho_{\mr{liquid}}$ and $\rho_{\mr{particle}}$ are the densities of solvent and particles, respectively. Therefore we can write the ratio of inter-particle distance to particle radius as $r/R=2/\phi^{1/3}=2\,(\rho_{\mr{particle}}/w \, \rho_{\mr{liquid}})^{1/3}$.

For aligned ellipsoidal particles of semiaxes $a$, $b$ and $c$ (case II), we can consider instead that the volume, $v$, surrounding each particle in the dispersion is a confocal ellipsoidal volume with semiaxis lengths $A$, $B$ and $C$ around the particle, so that $v = n^{-1} = (4/3)\pi A B C$. The inter-particle distances between the oriented ellipsoids along each direction can then be taken as $r_1=2A$, $r_2=2B$ and $r_3=2C$. Given that the particle and volume $v$ are confocal ellipsoids, we can write $A=\sqrt{a^2+\delta}$, $B=\sqrt{b^2+\delta}$, $C=\sqrt{c^2+\delta}$ and reduce the problem to finding $\delta$ in order to obtain $r_1$, $r_2$ and $r_3$. In this case, for a known volume or mass fraction, the number density is given by $n=\phi/V_{\mr{ellipsoid}}$ or $n=(\rho_{\mr{liquid}}/\rho_{\mr{particle}})\, w /V_{\mr{ellipsoid}}$, where $V_{\mr{ellipsoid}}=(4/3)\pi a b c$, and the ratios $r_1/a$, $r_2/b$ and $r_3/c$ now depend on the particle dimensions, $a$, $b$ and $c$. It is important to consider that for anisotropic particles, excluded volume and entropic effects can dominate at high concentrations, determining the ordering of particles in solution (e.g., liquid crystals).  
 
For the ellipsoidal volume model (case II) and for our lipid-coated HOPG micro-flakes we calculate that a volume fraction $\phi \leq 2.5 \,\mr{vol}\%$ (or a corresponding weight fraction $w \leq 1 \,\mr{wt}\%$) is required to guarantee that the inter-particle distance along each direction is at least 4 times the particle half-size along that direction ($r_1 \geq 4a$, $r_2 \geq 4b$, $r_3 \geq 4c$). We assume that, at such concentrations, dipole-dipole interactions can be neglected and the models presented in this manuscript can be applied. For higher particle concentrations, interactions become important and cannot be neglected. If we consider the spherical model (case I) instead, the limiting maximum concentration would be more generous, $\phi \leq 12 \,\mr{vol}\%$ ($w \leq 5 \,\mr{wt}\%$) to guarantee that $r \geq 4R$ for negligible interactions. 

%Relatedly, AC electric fields have recently been employed to orient and control the alignment of graphene oxide (GO) platelets (3.2 microns lateral size, monolayers) in dilute dispersions at concentrations below (but not far from) that of the nematic liquid crystal (LC) phase $\phi \leq 0.1 \,\mr{vol}\%$, to induce birefringence, a large optical Kerr coefficient and to generate optical switching for the fabrication of electro-optic display devices \cite{grapheneOxideMonolayersElectrAlign2014}. The authors show that control of GO particle orientation with AC electric fields is only efficient for dilute dispersions with negligible particle interactions. At higher particle concentrations, the transition to nematic LC phase takes place, governed mostly by excluded-volume effects, and the response of the GO-LC to the AC field is weak and increasingly reduced with density. 

\section{Depolarisation factors for ellipsoids} \label{App:depolarisationFactors}

The geometrical depolarization factors that account for shape anisotropy and for the reduction in field inside the particle due to its polarization, are given by the following expressions for the outer and inner ellipsoids, respectively:
{\setlength{\mathindent}{0cm}
	\begin{equation}\label{eqn:L1k}
	\small L_{\footnotesize 1k}=\frac{\small a_1 b_1 c_1}{2}\int^{\infty}_{0}\frac{\mr{d}s}{\left(s+p_{1k}^2 \right) \sqrt{\left(s+a_1^2 \right) \left(s+b_1^2 \right) \left(s+c_1^2 \right)} } \, ,
	\end{equation}
	\begin{equation}\label{eqn:L2k}
	\small L_{\footnotesize 2k}=\frac{\small a_2 b_2 c_2}{2}\int^{\infty}_{0}\frac{\mr{d}s}{\left(s+p_{2k}^2 \right) \sqrt{\left(s+a_2^2 \right) \left(s+b_2^2 \right) \left(s+c_2^2 \right)}} \, ,
	\end{equation}
	}
\normalsize \noindent with $p_{1k}=a_1,\,b_1,\,c_1$ and $p_{2k}=a_2,\,b_2,\,c_2$ for $k=x,\,y,\,z$, respectively. The following relations hold: \small{\hbox{$\sum_{k=x,y,z}L_{1k}=1$}} \normalsize and \small{$\sum_{k=x,y,z}L_{2k}=1$}. \normalsize Analytical expressions can be found for $L_{1k}$ and $L_{2k}$ in the case of oblate or prolate ellipsoids with two equal semi-axes \cite{TBJonesBook}. Note that for the particular case of a sphere, we have that $L_{1x}=L_{1y}=L_{1z}=1/3$, so that the effective polarization factor for a spherical uncoated particle with homogeneous complex permittivity $\epsilon_{2}$ [instead of $\epsilon'_{2k}$ in Eqns.(\ref{eqn:peff}) and (\ref{eqn:Kk})] would be the well-known Clausius-Mossotti factor \cite{ClausiusMossotti1,ClausiusMossotti2}.

\section{\textbf{Very-thin-layer approximation}} \label{App:VTLA}

When the layer is very thin, a very-thin-layer approximation (VTLA) can be made \cite{T3}. In this case, Eqn. (\ref{eqn:equivalentPermittivity}) is rearranged in terms of $(L_{2k}/\nu-L_{1k})$. In these terms, $L_{2k}$ is expressed in terms of $a_1$, $b_1$, $c_1$ and $\delta$, and a change of variable from $s$ to $\xi=s-\delta$ is used for the $L_{2k}$ integral. The trapezoidal rule is then used to approximate the resulting integral so that $(L_{2k}/\nu-L_{1k}) \approx t_k/p_{1k}$, with $t_k=t_x,\,t_y,\,t_z$ and $p_{1k}=a_1,\,b_1,\,c_1$ for $k=x,\,y,\,z$. The approximation $\nu \approx 1$ is also used. For an ellipsoid with $a_2>b_2>c_2$, the VTLA applies when the stringent condition $\delta \ll c_2^2$ is fulfilled, where $\delta \approx 2t_x a_2 \approx 2t_y b_2 \approx 2t_z c_2$. This results in an equivalent anisotropic complex permittivity for the layered ellipsoid \cite{T3}:
\begin{equation}\label{eqn:equivalentPermittivityTLA}
	\epsilon'_{2k} \approx 
		\epsilon_2 \cdot 
		\frac{\epsilon_{3k}+\left(\epsilon_{3k}-\epsilon_2 \right) \frac{t_k}{p_{1k}}}
		{\epsilon_2+\left(\epsilon_{3k}-\epsilon_2 \right) \frac{t_k}{p_{1k}}} \, .
\end{equation}
For a very thin layer, the depolarisation factors of the inner and outer ellipsoids can be approximated as equal ($L_{1k}\approx L_{2k}$) to write:
\begin{equation}\label{eqn:KkTLA}
	K_k \approx \frac{\epsilon'_{2k}-\epsilon_1}{\epsilon_1+(\epsilon'_{2k}-\epsilon_1)L_{2k}}\, .
\end{equation}
Within the VTLA, the equivalent permittivity along any of the principal axis directions depends on the ratio of the layer thickness to particle half-size along that direction ($t_k/p_{1k}$) and is independent of the layer thickness along the other directions. The VTLA has been employed to describe the dielectrophoresis and electro-rotation of biological cells (with thin lipid membranes) \cite{AsamiEllipsoidShell,T9} and the electro-orientation of erythrocytes \cite{T3}.

\section{\textbf{Dependence of $\epsilon'_{2k}$ and $K_k$ on the layer thickness along $k$}} \label{App:myApprox}

In this section, we show that for thin layers and moderate aspect ratios ($\nu\approx1$; for our experimental parameters $\nu \geq 0.8$), $K_x$ depends mainly on $t_x$ and $K_z$ depends mainly on $t_z$. Therefore, to calculate the maximum torque (proportional to $\mr{Re}(K_x-K_z)$) for a desired uniform shell thickness, $t_\mr{shell}$, a \emph{dual-ellipsoid} model can be used that calculates $K_x$ for an ellipsoid with $t_x=t_\mr{shell}$ and $K_z$ for a different ellipsoid with $t_z=t_\mr{shell}$.

In the expression for the equivalent complex permittivity for the layered ellipsoid [$\epsilon'_{2k}$ in Eqn. (\ref{eqn:equivalentPermittivity})], we first focus on the terms $\left( L_{2k}-\nu L_{1k} \right)$ that appear in the numerator and denominator. These terms reflect the difference in geometry between the inner and outer ellipsoids. In the expression for $L_{1k}$ [Eqn. (\ref{eqn:L1k})], we can substitute for $a_1^2=a_2^2+\delta$, $b_1^2=b_2^2+\delta$, $c_1^2=c_2^2+\delta$ and $p_{1k}^2=p_{2k}^2+\delta$ in order to express everything in terms of the inner ellipsoid semi-axes ($a_2$, $b_2$, $c_2$) and $\delta$. Following this, with a change of variable in the integral from $s$ to $\xi=s+\delta$, $L_{1k}$ becomes:
{\setlength{\mathindent}{0cm}
	\begin{equation}\label{eqn:L1kbis}
	\small L_{\footnotesize 1k}=\frac{\small a_1 b_1 c_1}{2}\int^{\infty}_{\delta}\frac{\mr{d}\xi}{\left(\xi+p_{2k}^2 \right) \sqrt{\left(\xi+a_2^2 \right) \left(\xi+b_2^2 \right) \left(\xi+c_2^2 \right)} } \, ,
	\end{equation}
	}
\normalsize \noindent and therefore $\left( L_{2k}-\nu L_{1k} \right)$ can be written as:
\begin{multline}\label{eqn:diffPolFactors}
	\left( L_{2k}-\nu L_{1k} \right) \equiv L_{k,\mr{diff}} = \\ 
	  \frac{\small a_2 b_2 c_2}{2}\int^{\delta}_{0}\frac{\mr{d}\xi}{\left(\xi+p_{2k}^2 \right) \sqrt{\left(\xi+a_2^2 \right) \left(\xi+b_2^2 \right) \left(\xi+c_2^2 \right)} }  \, .
\end{multline} 
These terms depend only on the inner ellipsoid semi-axes and on $\delta$ via the integral limit. For a confocal layered ellipsoid, given that $\delta=a_1^2-a_2^2=b_1^2-b_2^2=c_1^2-c_2^2$ and $a_1=a_2+t_x$, $b_1=b_2+t_y$, $b_1=b_2+t_z$, we have that $\delta=t_x^2+2t_x a_2=t_y^2+2t_y b_2=t_z^2+2t_z c_2=\mr{const}$. $\delta$ can be chosen in order to have the desired layer thickness along a given direction $k$. For instance, for a desired layer thickness $t_\mr{shell}$, a value $\delta=t_\mr{shell}^2+2t_\mr{shell} a_2$ will result in a layer thickness equal to $t_\mr{shell}$ along the $x$ direction and layer thicknesses along $y$ and $z$ different from $t_\mr{shell}$. With the choice of $\delta$ for the desired thickness along the $k$ direction we have that $\left( L_{2k}-\nu L_{1k} \right)$ effectively depends only on the layer thickness along $k$. This dependence is approximately linear for thin layers.

In Eqn. (\ref{eqn:equivalentPermittivity}), the additional $\nu$ term in the numerator [last term in $\left( L_{2k}-\nu L_{1k}+\nu \right)$] brings in an additional dependence for $\epsilon'_{2k}$ on the layer thicknesses along the directions different to $k$. The factor $L_{1k}$ in the denominator in Eqn.(\ref{eqn:Kk}) also brings in such dependence for $K_k$. $\nu$ and $L_{1k}$ are both geometrical factors. For ellipsoids with thin layers ($t_x\ll a_2$, $t_y\ll b_2$, $t_z\ll c_2$) and moderate aspect ratios that fulfill $\nu \approx 1$ it is a good approximation to consider that $K_k$ depends mainly on the layer thickness along the $k$ direction. 

\begin{figure}[ht]
	\centering
  \includegraphics[width=0.7\columnwidth]{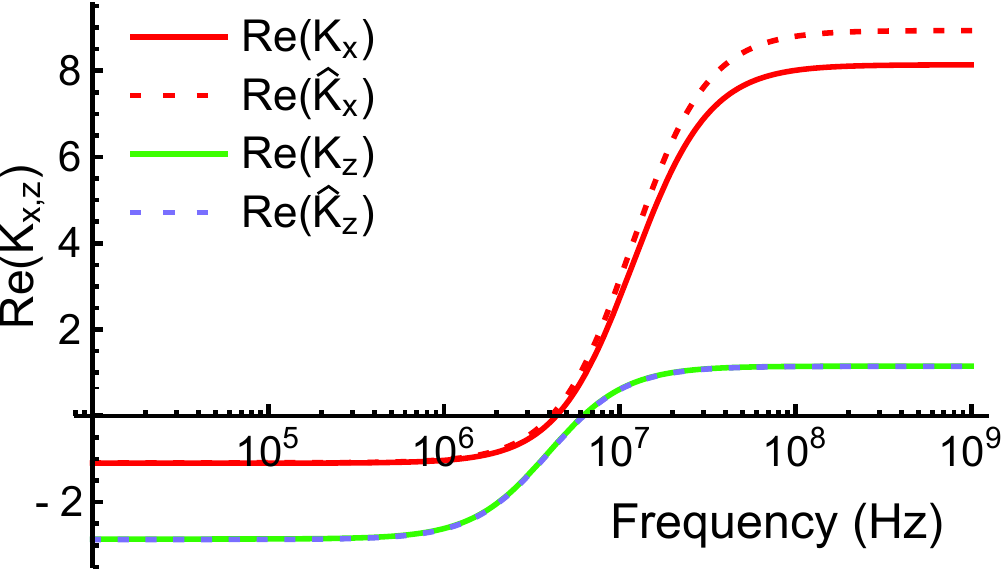}
  \caption{Calculated polarization factors versus frequency for layered ellipsoid with $a_2=3.4\,\mu\mr{m}$, $b_2=1.4\,\mu\mr{m}$ and $c_2=0.6\,\mu\mr{m}$. \textbf{Solid lines}: full calculations for $\mr{Re}(K_x)$ and $\mr{Re}(K_z)$ for two different ellipsoids with $t_x=15\,\mr{nm}$ and $t_z=15\,\mr{nm}$, respectively. \textbf{Dotted lines}: approximations $\mr{Re}(\hat{K}_x)$ and $\mr{Re}(\hat{K}_z)$ for the same $t_x$ and $t_z$. $\mr{Re}(K_x)$ and $\mr{Re}(\hat{K}_x)$ differ by at most $10\%$.}
  \label{fig:KxKzVsFreqApprox}
\end{figure}

To validate this assumption, we compare the full calculation for $K_k$ [Eqn.(\ref{eqn:Kk})] with an approximation, $\hat{K}_k$, that depends only on the desired thickness $t_\mr{shell}$ and for which $t_k = t_\mr{shell}$. This approximation uses an approximate $\hat{\nu} = a_2 b_2 c_2/(\hat{a}_1 \hat{b}_1 \hat{c}_1)$, where $\hat{a}_1=a_2+t_\mr{shell}$, $\hat{b}_1=b_2+t_\mr{shell}$ and $\hat{c}_1=c_2+t_\mr{shell}$, and also an approximate $\hat{L}_{1k}$:
{\setlength{\mathindent}{0cm}
	\begin{equation}\label{eqn:L1kApprox}
	\small \hat{L}_{\footnotesize 1k}=\frac{\small \hat{a}_1 \hat{b}_1 \hat{c}_1}{2}\int^{\infty}_{0}\frac{ds}{\left(s+\hat{p}_{1k}^2 \right) \sqrt{\left(s+\hat{a}_1^2 \right) \left(s+\hat{b}_1^2 \right) \left(s+\hat{c}_1^2 \right)}},
	\end{equation}}
\normalsize\noindent with $\hat{p}_{1k}=\hat{a}_1,\,\hat{b}_1,\,\hat{c}_1$ for $k=x,\,y,\,z$, respectively. In this way, the approximate effective polarization factors are calculated as:
\begin{equation}\label{eqn:KkApprox}
	\hat{K}_k = \frac{\hat{\epsilon}'_{2k}-\epsilon_1}{\epsilon_1+(\hat{\epsilon}'_{2k}-\epsilon_1)\hat{L}_{1k}}\, ,
\end{equation}
where the approximate equivalent permittivities are calculated as:
\begin{equation}\label{eqn:equivalentPermittivityApprox}
	\hat{\epsilon}'_{2k} = 
		\epsilon_2 \cdot 
		\frac{\epsilon_2+\left(\epsilon_{3k}-\epsilon_2 \right) \left( L_{k,\mr{diff}}+\hat{\nu} \right)}
		{\epsilon_2+\left(\epsilon_{3k}-\epsilon_2 \right) L_{k,\mr{diff}}} \, ,
\end{equation}
with $L_{k,\mr{diff}}$ defined in Eqn. (\ref{eqn:diffPolFactors}). The volume factor in the maximum torque [Eqn. (\ref{eqn:Tmax})] can also be calculated as $\frac{4}{3}\pi \hat{a}_1 \hat{b}_1 \hat{c}_1$. In this approximation, $\hat{K}_k$ depends only on the chosen shell thickness for the $k$ direction, $t_\mr{shell}$. Comparing the full calculation [$K_k$ in Eqn. (\ref{eqn:Kk})] with this approximation [$\hat{K}_k$ in Eqn. (\ref{eqn:equivalentPermittivityApprox})] for our experimental parameters, we find that the full calculation remains close to the approximation to within $\sim10-14\%$. An example is shown in Fig. \ref{fig:KxKzVsFreqApprox} for $\mr{Re}(K_x)$ and $\mr{Re}(K_z)$. This validates the assumption that the main dependence of $\mr{Re}(K_x)$ is on $t_x$, and the main dependence of $\mr{Re}(K_z)$ is on $t_z$ for thin layers and moderate aspect ratios ($\nu \geq 0.8$). Hence a \emph{dual-ellipsoid} calculation is proposed as an alternative to the one-ellipsoid model to improve predictions for the torque of uniformly-coated ellipsoids.

\section{Two-shell model} \label{App:theoryTwoShell}

\begin{figure}[ht]
	\centering
  \includegraphics[width=0.85\columnwidth]{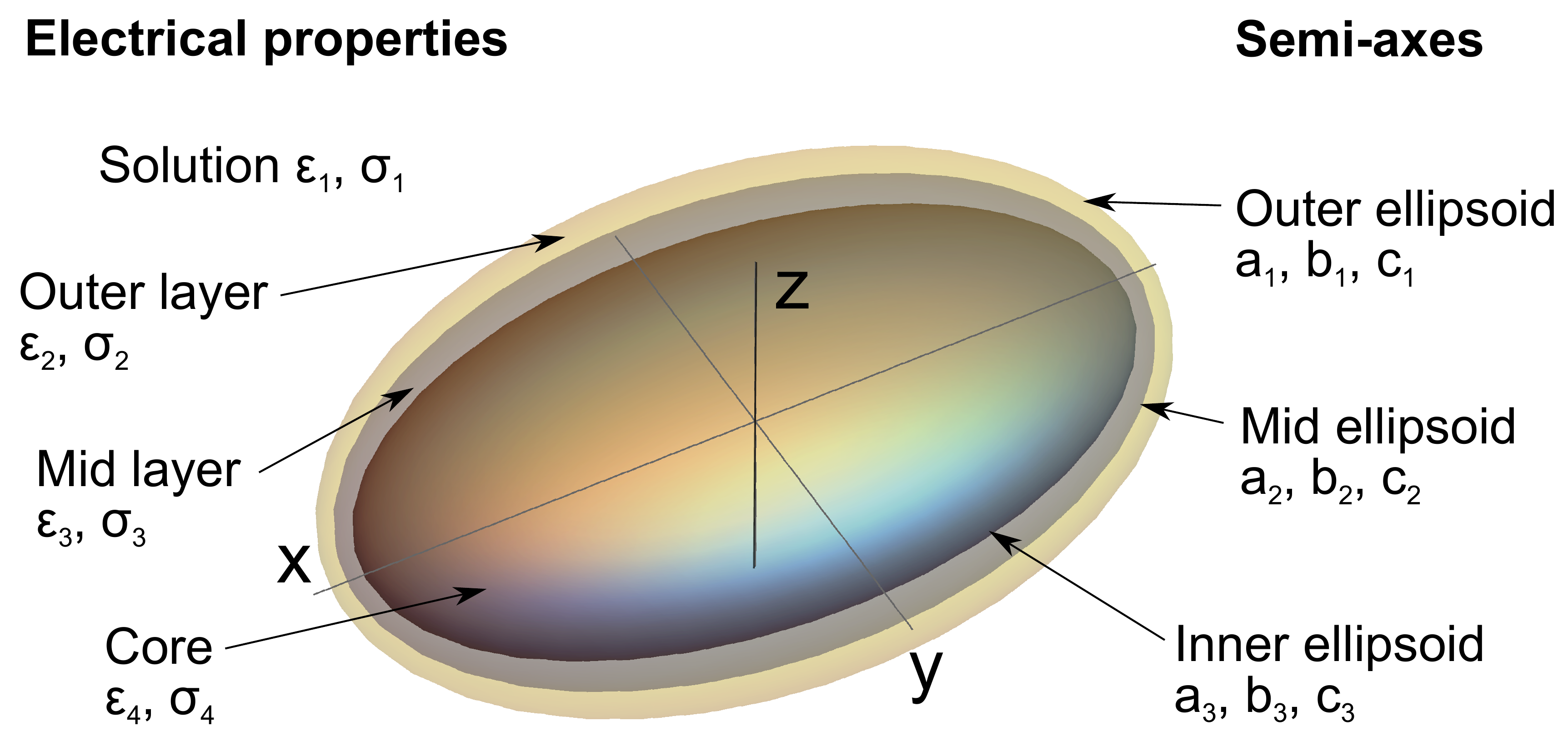}
  \caption{Schematic of ellipsoid with two thin shells in solution.}
  \label{fig:TwoLayerEllipsoid}
\end{figure}

We consider an ellipsoidal core (HOPG) with semi-axes $a_3,\,b_3,\,c_3$ (Fig. \ref{fig:TwoLayerEllipsoid}) and electrical properties given by $\epsilon_{4\parallel} = \varepsilon_{4\parallel}-\frac{\ii \sigma_{4\parallel}}{\omega \varepsilon_0}$ and $\epsilon_{4\perp} = \varepsilon_{4\perp}-\frac{\ii \sigma_{4\perp}}{\omega \varepsilon_0}$. The intermediate aqueous shell has electrical properties given by $\epsilon_3 = \varepsilon_3-\frac{\ii \sigma_3}{\omega \varepsilon_0}$, the outer shell (lipids) is described by $\epsilon_2 = \varepsilon_2-\frac{\ii \sigma_2}{\omega \varepsilon_0}$, and $\epsilon_1 = \varepsilon_1-\frac{\ii \sigma_1}{\omega \varepsilon_0}$ corresponds to the solution. The sizes of the intermediate ellipsoid (core and intermediate shell) and outer ellipsoid (core and two shells) are given by the semi-axes $a_2,\,b_2,\,c_2$ and $a_1,\,b_1,\,c_1$, respectively (see Fig. \ref{fig:TwoLayerEllipsoid}). For the three confocal ellipsoids we have $a_2=(a_3^2+\delta_2)^{1/2}$, $b_2=(b_3^2+\delta_2)^{1/2}$ and $c_2=(c_3^2+\delta_2)^{1/2}$ and $a_1=(a_3^2+\delta_2+\delta_1)^{1/2}$, $b_1=(b_3^2+\delta_2+\delta_1)^{1/2}$ and $c_1=(c_3^2+\delta_2+\delta_1)^{1/2}$, where parameters $\delta_1$ and $\delta_2$ determine the thickness of the outer and intermediate shells, respectively. The thickness of the outer lipid shell along each direction is given by $t_{\mr{lipids},x}=a_1-a_2$, $t_{\mr{lipids},y}=b_1-b_2$ and $t_{\mr{lipids},z}=c_1-c_2$. The thickness of the intermediate aqueous shell along each direction is $t_{\mr{aq},x}=a_2-a_3$, $t_{\mr{aq},y}=b_2-b_3$ and $t_{\mr{aq},z}=c_2-c_3$.

The only difference with respect to the one-shell model is the form of the effective complex polarization factor, $K_k$, of the two-shell ellipsoid, which is given by:
\begin{equation}\label{eqn:Kk_app}
	K_k = \frac{\epsilon'_{2k}-\epsilon_1}{\epsilon_1+(\epsilon'_{2k}-\epsilon_1)L_{1k}}\, .
\end{equation}
where $\epsilon'_{2k}$ is the equivalent complex permittivity of the two-shell ellipsoid, given by:
\begin{equation}\label{eqn:equivalentPermittivity_app}
	\epsilon'_{2k} = 
		\epsilon_2 \cdot 
		\frac{\epsilon_2+\left(\epsilon'_{3k}-\epsilon_2 \right) \left[ L_{2k}+\nu_2 \left(1-L_{1k}\right)\right]}
		{\epsilon_2+\left(\epsilon'_{3k}-\epsilon_2 \right) \left[ L_{2k}-\nu_2 L_{1k} \right]} \, .
\end{equation}
Here, $\nu_2=(a_2 b_2 c_2)/(a_1 b_1 c_1)$ and $L_{1k}$, $L_{2k}$ ($k=x,\,y,\,z$) are the geometrical depolarization factors of the outer and intermediate ellipsoids, respectively, defined as in Eqs.(\ref{eqn:L1k}) and (\ref{eqn:L2k}). $\epsilon'_{3k}$ is the equivalent complex permittivity of the intermediate ellipsoid, given by:
\begin{equation}\label{eqn:equivalentPermittivity_app}
	\epsilon'_{3k} = 
		\epsilon_3 \cdot 
		\frac{\epsilon_3+\left(\epsilon_{4k}-\epsilon_3 \right) \left[ L_{3k}+\nu_3 \left(1-L_{2k}\right)\right]}
		{\epsilon_3+\left(\epsilon_{4k}-\epsilon_3 \right) \left[ L_{3k}-\nu_3 L_{2k} \right]} \, ,
\end{equation}
where $\nu_3=(a_3 b_3 c_3)/(a_2 b_2 c_2)$ and $L_{3k}$ ($k=x,\,y,\,z$) are the depolarization factors of the inner ellipsoid, defined as:
{\setlength{\mathindent}{0cm}
	\begin{equation}\label{eqn:L3k}
	\small L_{\footnotesize 3k}=\frac{\small a_3 b_3 c_3}{2}\int^{\infty}_{0}\frac{ds}{\left(s+p_{3k}^2 \right) \sqrt{\left(s+a_3^2 \right) \left(s+b_3^2 \right) \left(s+c_3^2 \right)}},
	\end{equation} 
	}
\normalsize\noindent with $p_{3k}=a_3,\,b_3,\,c_3$ for $k=x,\,y,\,z$, respectively. $\epsilon_{4k}$ is the complex permittivity of the inner core so that \small$\epsilon_{4x} = \epsilon_{4y} = \varepsilon_{\mr{hopg},\parallel}-\ii \sigma_{\mr{hopg},\parallel}/(\omega \varepsilon_0)$ \normalsize and \small$\epsilon_{4z} = \varepsilon_{\mr{hopg},\perp}-\ii \sigma_{\mr{hopg},\perp}/(\omega \varepsilon_0)$ \normalsize for the in-plane and out-of-plane HOPG directions, respectively.

The two-shell \emph{dual-ellipsoid} model separates calculations of the effective complex polarization factors for the relevant $x$ and $z$ directions. A two-shell ellipsoid with the desired layer thicknesses $t_{\mr{aq},x}$ and $t_{\mr{lipids},x}$ is used to calculate $K_x(\omega)$ while a different two-shell ellipsoid with the desired $t_{\mr{aq},z}$ and $t_{\mr{lipids},z}$ is used to calculate $K_z(\omega)$. 

\begin{figure}[t]
	\centering
  \includegraphics[width=0.7\columnwidth]{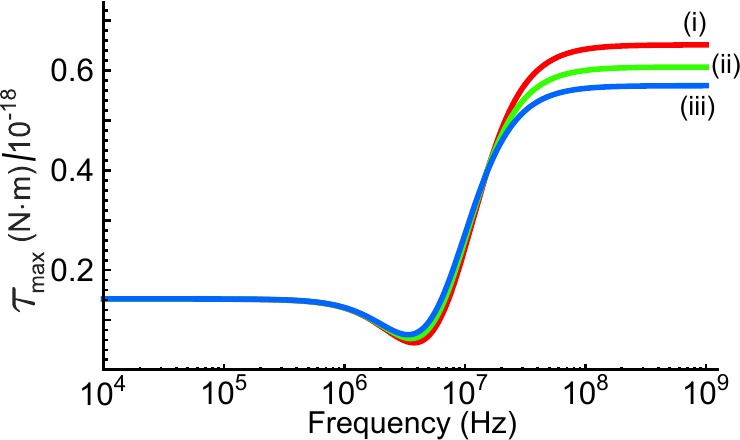}
  \caption{$\mathcal{T}_\mr{max}$ versus frequency calculated with the two-shell \emph{dual-ellipsoid} model for an HOPG core with $a_3=2.5\,\mu\mr{m}$, $b_3=1\,\mu\mr{m}$ and $c_3=0.5\,\,\mu\mr{m}$ and for: \textbf{(i)} no intermediate aqueous layer ($t_{\mr{aq},x}=t_{\mr{aq},z}=0$) and 10nm-thick lipid layer ($t_{\mr{lipids},x}=t_{\mr{lipids},z}=10\,\mr{nm}$) \textbf{(ii)} aqueous layer with $t_{\mr{aq},x}=t_{\mr{aq},z}=10\,\mr{nm}$ and lipid layer with $t_{\mr{lipids},x}=t_{\mr{lipids},z}=10\,\mr{nm}$, \textbf{(iii)} $t_{\mr{aq},x}=t_{\mr{aq},z}=20\,\mr{nm}$ and $t_{\mr{lipids},x}=t_{\mr{lipids},z}=10\,\mr{nm}$.}
  \label{fig:twoLayers}
\end{figure}

We now apply the two-shell \emph{dual-ellipsoid} model to predict the torque for lipid-coated HOPG micro-flakes with a thin intermediate aqueous solution layer between the lipid shell and the graphitic core. We use the following parameters: $E_0 = 1 \times 10^4\,\mr{V/m}$, $a_3=2.5\,\mu \mathrm{m}$, $b_3=1\,\mu \mathrm{m}$ and $c_3=0.5\,\mu \mathrm{m}$ for the HOPG core, $\varepsilon_1=\varepsilon_3=80$ and $\sigma_1=\sigma_3=0.2\,\mr{S/m}$ for the external aqueous solution and for the intermediate aqueous layer (both 20 mM NaCl), $\varepsilon_2=11$ and $\sigma_2=10^{-7}\,$S/m for the outer lipid shell, and $\varepsilon_{\mr{hopg},\parallel}=2.4$, $\sigma_{\mr{hopg},\parallel}=2.5 \times 10^6\,$S/m, $\varepsilon_{\mr{hopg},\perp}=1.75$ and $\sigma_{\mr{hopg},\perp}=250\,$S/m for the HOPG core. Fig. \ref{fig:twoLayers} shows the predicted torque versus frequency for an increasing thicknesss of the intermediate aqueous layer (0, 10, $20\,\mr{nm}$, curves (i) to (iii)). In this case, it is harder to provide intuitive explanations given the complex interplay of resistive, capacitive and polarisation effects of the various layers. An increasing intermediate-layer thickness seems to shift the main dispersion to slightly lower frequencies. The effect appears to be opposite to that of increasing the thickness of the lipid layer, pointing to the contribution of polarisation effects at the new lipid-water (and water-HOPG) interfaces. 

\section{Changing particle eccentricity while keeping a constant ratio $t_x/a_2$} \label{App:changeEccentrSuppl}

\begin{figure}[ht]
	\centering
  \includegraphics[width=0.75\columnwidth]{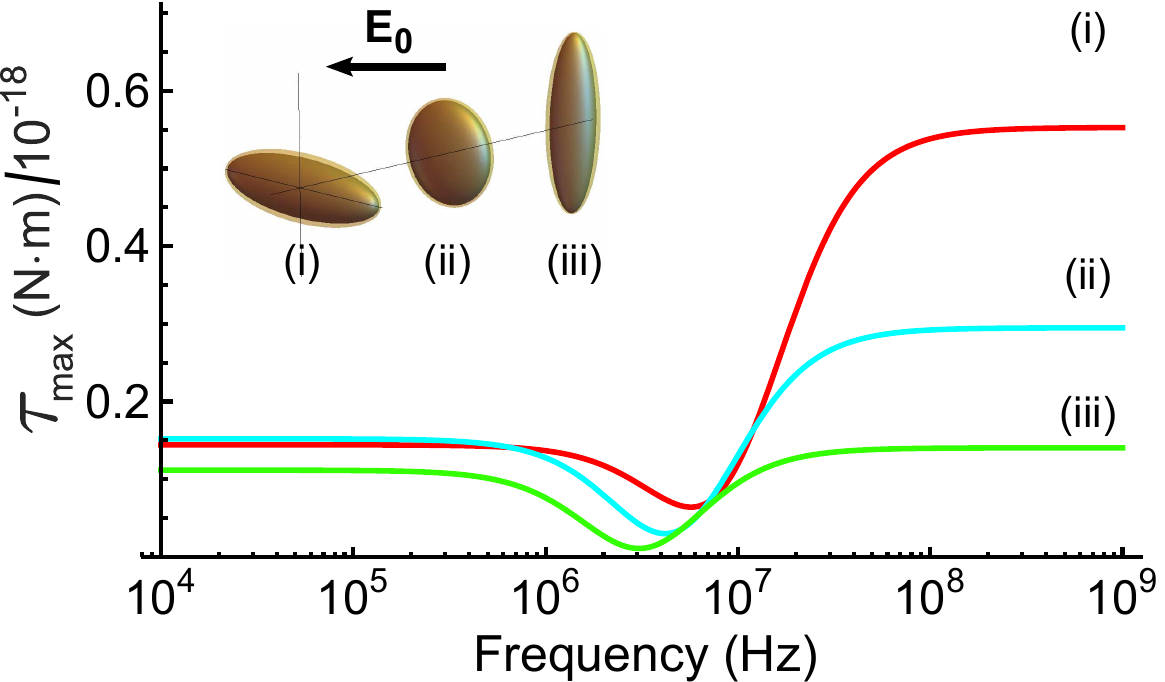}
  \caption{$\mathcal{T}_\mr{max}$ versus frequency calculated with the one-shell \emph{dual-ellipsoid} model for different ratios $a_2/b_2$ (particle eccentricity on the $x$-$y$ plane), for particles of equal volume and equal ratio ($t_x/a_2$) of lipid-layer thickness to particle size along $x$: (i) $a_2=2.5\,\mu\mr{m}$, $b_2=1\,\mu\mr{m}$, $c_2=0.5\,\,\mu\mr{m}$ and $t_\mr{shell}=16\,\mr{nm}$, (ii) $a_2=b_2=1.58\,\mu\mr{m}$, $c_2=0.5\,\,\mu\mr{m}$ and $t_\mr{shell}=10\,\mr{nm}$ and (iii) $a_2=1\,\mu\mr{m}$, $b_2=2.5\,\mu\mr{m}$, $c_2=0.5\,\,\mu\mr{m}$ and $t_\mr{shell}=6.3\,\mr{nm}$.}
  \label{fig:changeEccentrSuppl}
\end{figure}

Figure \ref{fig:changeEccentrSuppl} shows the effect of changing the eccentricity, $a_2/b_2$, of the particle on the $x$-$y$ plane while keeping a constant particle volume, a constant size along $z$ ($c_2=0.5\,\,\mu\mr{m}$) and a constant ratio ($t_x/a_2\approx0.006$) of lipid-layer thickness to particle size along $x$. The one-shell \emph{dual-ellipsoid} model is used. The curves shown correspond to: (i) $a_2=2.5\,\mu\mr{m}$, $b_2=1\,\mu\mr{m}$ and $t_\mr{shell}=16\,\mr{nm}$, (ii) $a_2=b_2=1.58\,\mu\mr{m}$ and $t_\mr{shell}=10\,\mr{nm}$ and (iii) $a_2=1\,\mu\mr{m}$, $b_2=2.5\,\mu\mr{m}$ and $t_\mr{shell}=6.3\,\mr{nm}$. The curves are very similar to those in Fig. \ref{fig:calcTvsFreq_Eccentricity}, showing that changing the size of the particle along the in-plane $x$ direction is most relevant when changing particle eccentricity, compared to changes in the ratio $t_x/a_2$.

\section{Modifying the in-plane conductivity of the particle} \label{App:changeHOPGcond}
 
\begin{figure}[ht]
	\centering
  \includegraphics[width=0.75\columnwidth]{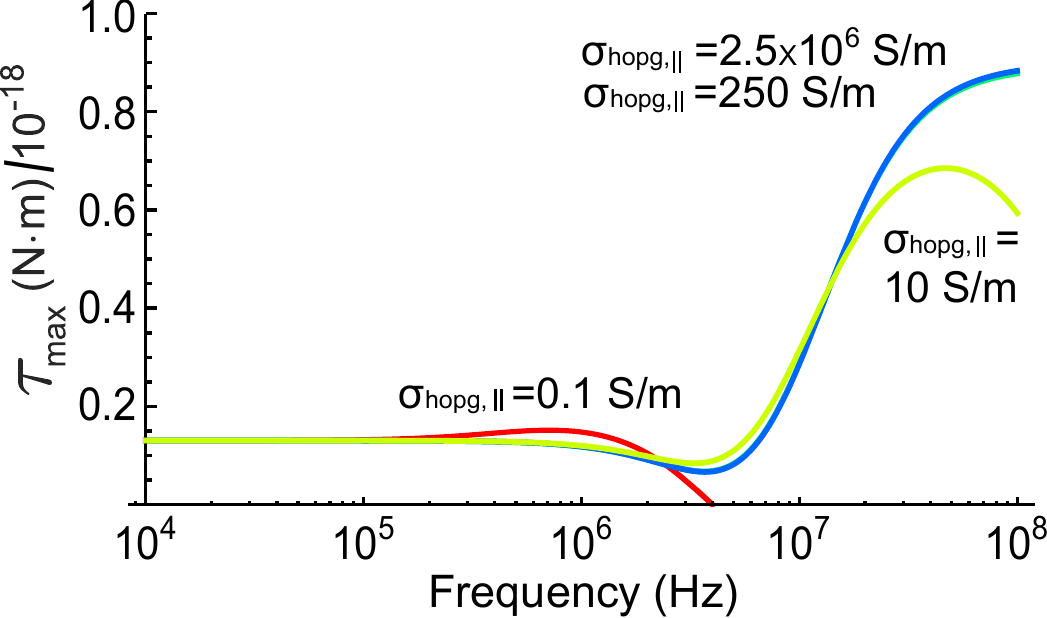}
  \caption{$\mathcal{T}_\mr{max}$ versus frequency for different hypothetical in-plane conductivities of the particle, $\sigma_{\mr{hopg},\parallel}$: 0.1 S/m, 10 S/m, 250 S/m and $2.5 \times 10^6\,\mr{S/m}$. All curves are calculated using the one-shell \emph{dual-ellipsoid} model with $a_2=3\,\mu \mathrm{m}$, $b_2=0.83\,\mu \mathrm{m}$, $c_2=0.5\,\mu \mathrm{m}$ and $t_\mr{shell}=10\,\mr{nm}$.}
  \label{fig:changeHOPGconduct}
\end{figure}

Figure \ref{fig:changeHOPGconduct} shows $\mathcal{T}_\mr{max}$ versus frequency calculated with the one-shell \emph{dual-ellipsoid} model for different hypothetical values of the in-plane conductivity of the particle: 0.1 S/m, 10 S/m, 250 S/m and $2.5 \times 10^6\,\mr{S/m}$. The actual in-plane and out-of plane conductivities of HOPG are $\sigma_{\mr{hopg},\parallel}=2.5 \times 10^6\,$S/m and $\sigma_{\mr{hopg},\perp}=250\,$S/m, respectively \cite{HOPGdielectricProperties}. Changing the in-plane conductivity of the particle has no effect unless this value becomes comparable to or lower than the conductivity of the solution. 

\section{Electric field at the sample region} \label{App:EfieldWires}

\begin{figure}[ht]
	\centering
  \includegraphics[width=0.9\columnwidth]{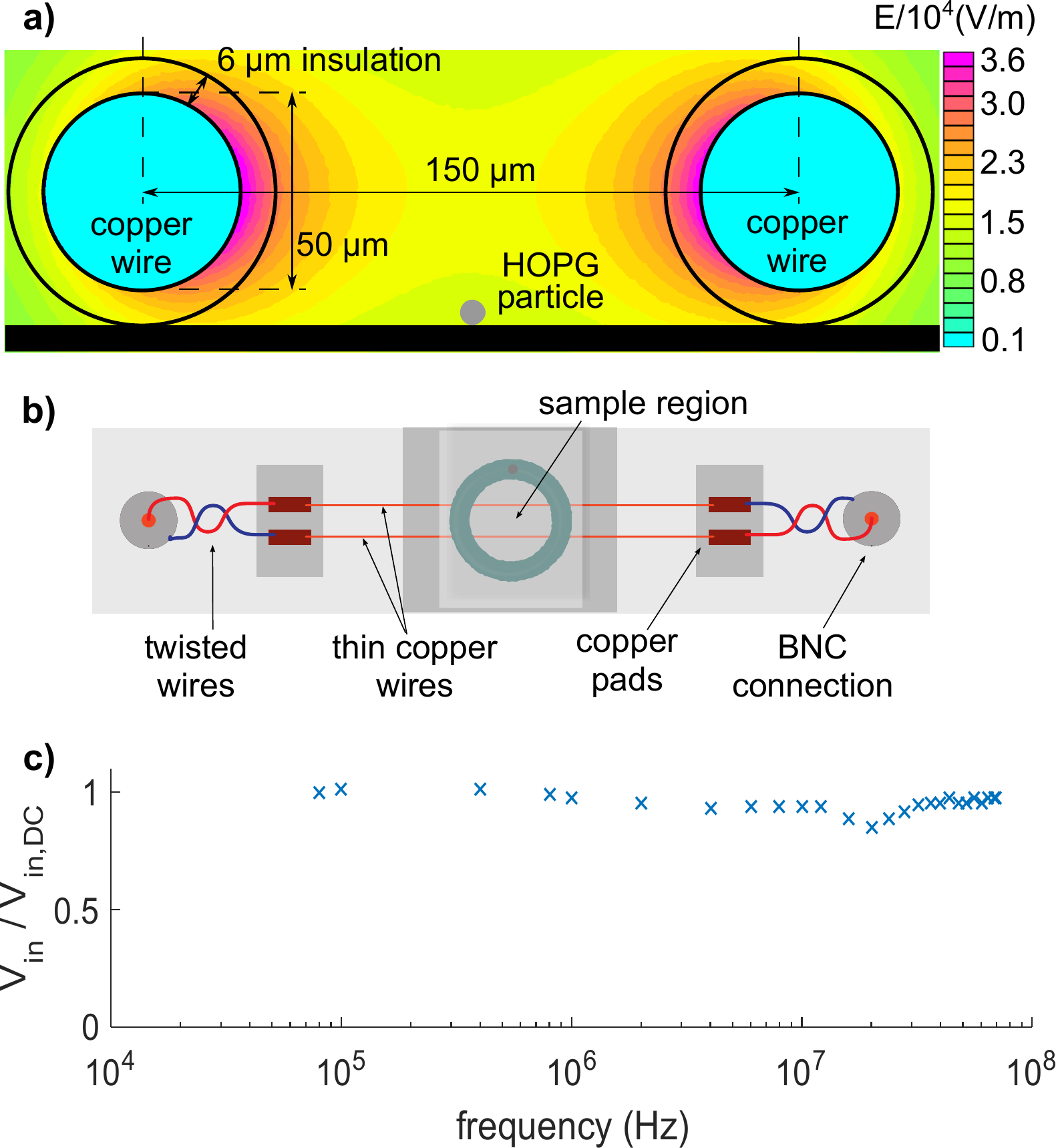}
  \caption{\textbf{a)} Map of the electric field amplitude between two parallel, $50\,\mu \mathrm{m}$-diameter wires with a $6\,\mu \mathrm{m}$-thick insulation layer, placed on top of a glass coverslip at a centre-to-centre distance of $150\,\mu\mr{m}$, when an AC voltage difference of amplitude 2.3 V is applied. \textbf{b)} Schematic of electrical connections generating a MHz AC electric field at the sample region between the two thin wire electrodes. \textbf{c)} Measured amplified input voltage amplitude, $V_\mr{in}$, into the wire electrodes as a function of frequency. All values are normalised by the input voltage amplitude at DC frequency, $V_\mr{in,DC}$.}
  \label{fig:EfieldWires}
\end{figure}

Figure \ref{fig:EfieldWires}a shows a map of the theoretical electric field amplitude between the sample electrodes when an AC voltage difference with a 2.3 V amplitude is applied between them. The map predicts an electric field amplitude in the region where the micro-particles are located, near the glass substrate, of $E_0 \approx 1.5 \times 10^4\,\mr{V/m}$. This corresponds to the ideal case of perfect impedance matching in all connections between the voltage source and the electrode wires (Fig. \ref{fig:EfieldWires}b).

An AC voltage source was connected to the electrodes through a series of connecting wires, as shown schematically in Fig. \ref{fig:EfieldWires}b. BNC coaxial cables were connected to intermediate twisted wires and then to the thin wire electrodes via a soldering connection onto copper pads. The BNC cables had $50\,\Omega$ impedance, and the twisted and thin electrode wires had approximately $100\,\Omega$ impedance \cite{magnetoElectrOrientHOPG}. The voltage from an RF signal generator and amplifier (specified for transfer to a $50\,\Omega$-impedance load) was fed to the input BNC connection on one end whereas the output BNC on the other was connected to a $50\,\Omega$ terminator to minimize signal reflections. Fig. \ref{fig:EfieldWires}c shows $V_{in}/V_{in,DC}$, the measured input voltage amplitude after amplification as a function of frequency, with all values normalised to the input voltage amplitude at DC ($V_{in,DC}$). The figure reveals a small variation with frequency of the voltage amplitude fed to the electrode wires that is used to correct the experimental data for the measured torque (section \ref{sec:results:compareToExperim}).

At the high MHz frequencies employed in experiments, impedance matching critically determines the actual AC voltage amplitude between the sample electrodes. In contrast to DC or low-frequency signals, the voltage wave generated by the signal generator will reflect back if it encounters connection interfaces with mismatched impedances. In our set-up, some impedance mismatch was unavoidable given that the impedance of the sample electrodes (parallel-wire transmission line) could not actually be matched to $50\,\Omega$ due to geometrical constraints, and was instead $\sim100\,\Omega$. Particular care was therefore placed in tapering connecting wire distances in an effort to avoid abrupt changes in impedance. This resulted in reasonably good actual impedance matching, evidenced by the relatively low transmission losses measured (mostly below $10\,\%$ and up to $20\,\%$ specifically around 20 MHz and 70 MHz) \cite{magnetoElectrOrientHOPG}.   

Measuring the actual electric field between the $50\,\mu\mr{m}$-diameter sample electrodes is not an easy task given the very small ($100\,\mu\mr{m}$-wide) gap between them and the fact that any probes introduced will modify the field between the electrodes. In order to fully characterise and avoid signal losses and quantify the exact voltage drop between micron-sized electrodes, a vector network analyser (VNA) can be used to analyse the voltage amplitude and phase in the relevant nodes of the electrical network. This can be combined with finite-element simulations and requires a non-trivial effort and the availability of an expensive VNA. Examples of such analysis for the design of impedance-matched circuits of micro-electrodes for biosensing at MHz frequencies can be found in the literature \cite{impedanceAnalysisVNA_1,impedanceAnalysisVNA_2}. 

% Note that this figure is part of the next section but I did not find a way to place it properly other than this:

\section{Experimental data for all particles} \label{App:allTorqueData}

\begin{figure*}[t]
	\centering
	\includegraphics[width=0.95\textwidth]{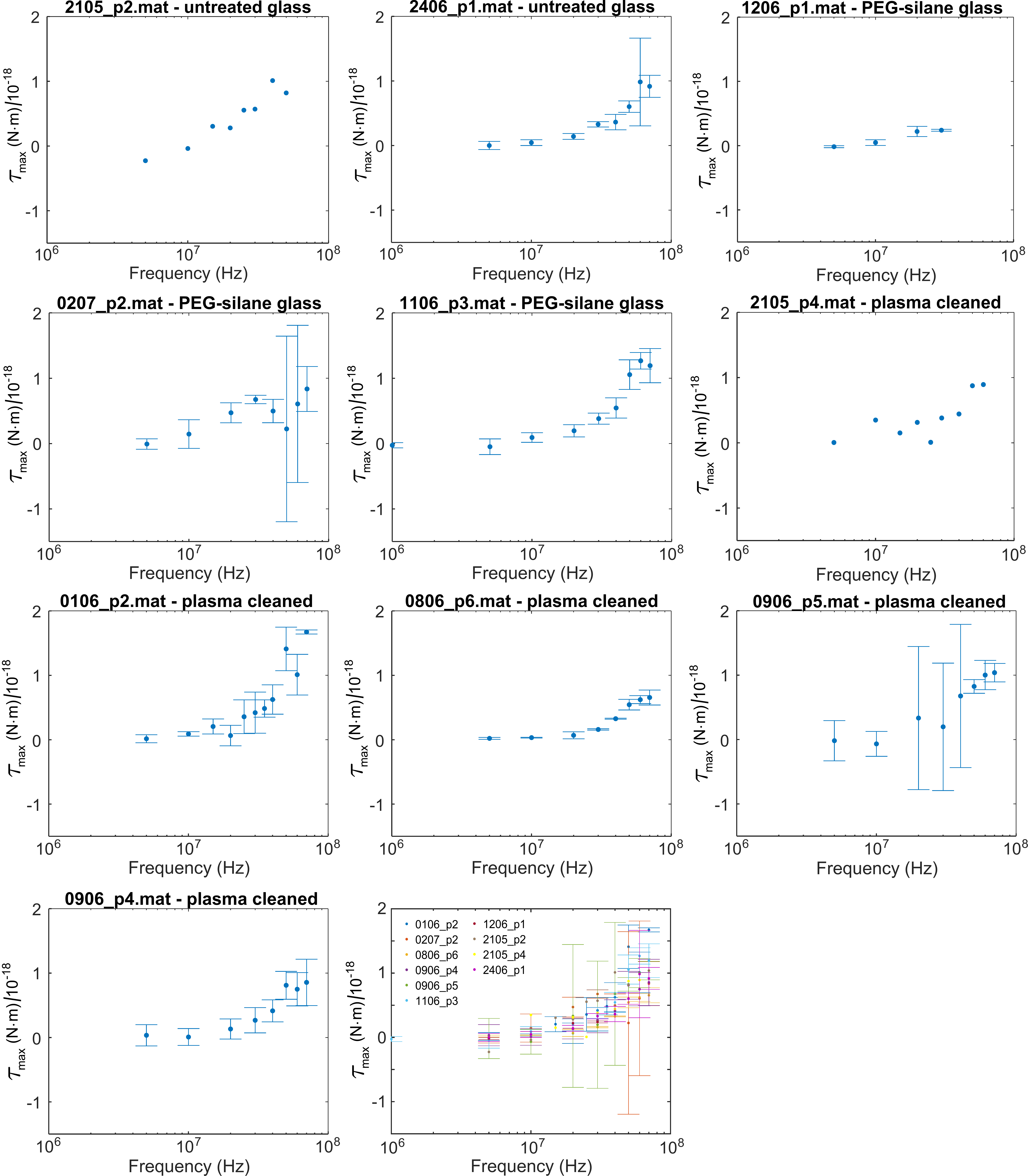}
  \caption{Measured maximum electric torque, $\mathcal{T}_\mr{max}$, as a function of frequency for 10 HOPG micro-particles on untreated glass, PEG-silane passivated glass and plasma-cleaned glass substrates, as indicated by each plot title. The last plot shows measurements for all particles together.}
	\label{fig:allParticlesData}
\end{figure*}

The measured maximum electric torque acting on 10 different individual lipid-coated HOPG micro-flakes submerged in 20 mM NaCl aqueous solution is shown in Fig. \ref{fig:allParticlesData}. The data corresponds to measurements with 2 micro-flakes on untreated glass, 3 micro-flakes on PEG-silane passivated glass and 5 HOPG micro-flakes on plasma-cleaned glass substrates. The error bars in the plots correspond to the standard deviation of the various measurements (2-5) performed at each frequency (no error bar is plotted if there is only one measurement). Having subtracted the trapping effect of the substrate to obtain the maximum electric torque (section \ref{sec:results:compareToExperim}), no significant differences are observed between the measurements for the different substrates.
%\begin{figure}[ht]
	%\centering
	%\includegraphics[width=0.85\columnwidth]{figures/allParticlesLog.pdf}
  %\caption{Measured $\mathcal{T}_\mr{max}$ versus frequency for all 10 individual HOPG micro-particles.}
	%\label{fig:torqueDataAll}
%\end{figure}

%\FloatBarrier

\section*{References}

\bibliography{HOPGelectroOrientBiblio_Elsv}

\end{document}